\documentclass[aps,twocolumn,showpacs]{revtex4}
\usepackage{graphicx}
\usepackage{dcolumn}
\usepackage{amsmath}
\usepackage{enumerate}

\begin{document}

\title{Uncertainties in nuclear transition matrix elements for $\beta^{+}\beta ^{+}$ and 
$\varepsilon \beta^{+}$ modes of neutrinoless positron double-$\beta $ decay within PHFB model}

\author{P. K. Rath$^{1}$, R. Chandra$^{2}$, K. Chaturvedi$^{3}$, 
P. Lohani$^{1}$, P. K. Raina$^{4}$ and J. G. Hirsch$^{5}$}

\affiliation{
$^{1}$Department of Physics, University of Lucknow, Lucknow-226007, India\\
$^{2}$Department of Applied Physics, Babasaheb Bhimrao Ambedkar University, 
Lucknow-226025, India\\
$^{3}$Department of Physics, Bundelkhand University, Jhansi-284128, India\\
$^{4}$Department of Physics, Indian Institute of Technology, Ropar,
Rupnagar - 140001, Punjab, India\\
$^{5}$Instituto de Ciencias Nucleares, Universidad Nacional Aut\'{o}noma de
M\'{e}xico, 04510 M\'{e}xico, D.F., M\'{e}xico}
\date{\today}

\begin{abstract}
Uncertainties in the nuclear transition matrix elements 
$M^{\left( 0\nu \right) }$ and  $M^{\left( 0N \right) }$ of the double-positron 
emission $(\beta^{+}\beta ^{+})_{0\nu }$ and electron-positron conversion 
$(\varepsilon \beta^{+})_{0\nu }$ modes due to the exchange of light and heavy 
Majorana neutrinos, respectively, are calculated for $^{96}$Ru, $^{102}$Pd, 
$^{106}$Cd, $^{124}$Xe, $^{130}$Ba and  $^{156}$Dy isotopes by employing the 
PHFB model with four different parameterization of the pairing plus multipolar 
two-body interactions and three different parameterizations of the Jastrow 
short range correlations. In all cases but for $^{130}$Ba, 
the uncertainties are smaller than 14\%  for light Majorana neutrino exchange 
and 35\%  for the exchange of a heavy Majorana neutrino.

\end{abstract}


\maketitle

\section{INTRODUCTION}

The Majorana nature of the neutrinos could be immediately established by 
confirming the possible occurrence of any one out of four experimentally
distinguishable modes of lepton number violating neutrinoless double beta $%
\left( \beta \beta \right) _{0\nu }$ decay, namely the double-electron
emission $(\beta ^{-}\beta ^{-})_{0\nu }$, double-positron emission $(\beta
^{+}\beta ^{+})_{0\nu }$, electron-positron conversion $(\varepsilon \beta
^{+})_{0\nu }$ and double-electron capture $(\varepsilon \varepsilon )_{0\nu
}$. The latter
three modes are energetically competing and we shall refer to them as $%
\left( e^{+}\beta \beta \right) _{0\nu }$ decay. The kinetic energy release
in the $(\varepsilon \varepsilon )_{0\nu }$ mode is the largest. However,
the conservation of energy-momentum requires the emission of an additional
particle in the $(\varepsilon \varepsilon )_{0\nu }$ mode. 
The absorption of atomic electrons from the $K$-shell is forbidden for 
the $0^{+}\rightarrow 0^{+}$ transition due to the emission of  one real photon. 
Consequently, various processes such as internal pair production, internal conversion,
emission of two photons, $L$-capture etc. \cite{doi93} have to be considered.
The decay rates of the above mentioned processes have to be 
calculated at least by the third order perturbation theory and are suppressed by a
factor of the order of 10$^{-4}$ in comparison to the $(\varepsilon \beta
^{+})_{0\nu }$ mode. Hence, the experimental as well as theoretical studies of 
$(e^{+}\beta \beta )_{0\nu }$ decay had been mostly restricted to 
$(\beta ^{+}\beta^{+})_{0\nu }$ and $(\varepsilon \beta ^{+})_{0\nu }$ 
modes only.

The idea behind the resonant enhancement of $(\varepsilon \varepsilon
)_{0\nu }$ mode \cite{wint55,volo82,verg83,bern83}, has been recently 
reanalyzed \cite{sujk04,luka06} and it has been shown that there will be
resonant enhancement of the $(\varepsilon \varepsilon )_{0\nu }$ mode upto a
factor of 10$^{6}$ provided the nuclear levels in the
parent and daughter nuclei are almost degenerate i.e. $Q-(E_{2P}-E_{2S})$ $%
\sim 1$ $keV$, where the energy difference is for atomic levels. Subsequently, 
detailed theoretical studies on the resonant enhancement of 
$(\varepsilon \varepsilon)_{0\nu }$ mode have also been performed \cite{kriv11,verg11}.
In the mean time, experimental studies on resonance enhancement of $(\varepsilon \varepsilon
)_{0\nu }$ mode in $^{74}$Ge \cite{bara07,frek11}, $^{96}$Ru \cite{bell09a}, 
$^{106}$Cd \cite{rukh10,bell12}, $^{112}$Sn \cite{bara08,daws08a,daws08b,kidd08,bara09}, 
$^{136}$Ce \cite{bell09b} and $^{180}$W \cite{bell11} isotopes 
have already been carried out and the study of this $
(\varepsilon \varepsilon )_{0\nu }$ mode is emerging as an interesting possibility 
for the investigation of $\left( \beta \beta \right) _{0\nu }$ decay.

In addition to establishing the Dirac or Majorana nature of neutrinos, the
observation of $\left( \beta \beta \right) _{0\nu }$ decay can also
ascertain the role of various mechanisms in different gauge theoretical
models \cite{klap06}. The study of $\left( \beta \beta \right)
_{0\nu }$ decay can clarify a number of issues, such as the origin of
the neutrino mass, their absolute scale as well as hierarchy, 
and possible CP violation in the leptonic sector. 
The $\left( \beta \beta \right) _{0\nu }$ and 
$\left( e^{+}\beta \beta \right) _{0\nu }$ decay modes can provide us with 
similar but complementary information. 
The observation of $(e^{+}\beta \beta )_{0\nu }$ decay modes would
be helpful in determining the presence of mass mechanism or 
right handed currents \cite{hirs94}. The varied scope and far reaching nature of
the experimental and theoretical studies on the $\left( \beta \beta\right) _{0\nu }$ decay 
have been recently reviewed by Avignone \textit{et al.} \cite{avig08}, 
Vergados \textit{et al.} \cite{verg12} and Faessler \textit{et al.} \cite{faes12}

The nuclear $\beta \beta $ decay proceeds through strongly suppressed channels 
which are very sensitive to details of the wave functions of
the parent, intermediate and daughter nuclei. Hence, the calculations of non-collective 
nuclear $\beta \beta $ decay related observables are quite challenging.
In any nuclear model, there are three basic ingredients, namely the model space, 
the single particle energies (SPEs) and the effective two body interactions. Usually, 
these are chosen on the basis of practical considerations. 
While all models are able to reproduce most of the observed 
$\left( \beta \beta \right) _{2\nu }$ decay half lives by adjusting
free parameters in the model, different predictions are obtained for other observables, 
like the $\left( \beta \beta \right) _{0\nu }$ decay half lives,
due to the inherent freedom in choosing the basic ingredients of the model.

A variety of nuclear models is currently employed in this endeavor. 
Large scale shell model calculations are quite successful 
\cite{stma,caur08,horo10}, but highly limited in the description of
medium and heavy mass nuclei. The most popular and successful model is 
the Quasiparticle Random Phase 
Approximation (QRPA) and its extensions \cite{suho98,faes98}.
The inclusion of nuclear deformation has also been carried out in the deformed 
QRPA \cite{simk04,fang11}, the Projected Hartree-Fock-Bogoliubov (PHFB) 
\cite{rath09,rath10a,rath11}, the pseudo-SU(3) \cite{jghi95}, the 
Interacting Boson Model (IBM) \cite{bare09}, and the Energy Density Functional (EDF) 
\cite{rodr10} approaches.
In the study of both $\left(
\beta \beta \right) _{2\nu }$ and $\left( \beta \beta \right) _{0\nu }$
decay modes, the renormalized value of axial vector coupling constant $g_{A}$
is a major source of uncertainty. In the $\left( \beta \beta \right) _{0\nu }$
decay, the role of pseudoscalar and weak magnetism terms \cite{simk99,verg02} 
is crucial, and the finite size of nucleons (FNS) and short range
correlations (SRC) play a decisive role vis-a-vis the radial evolution of nuclear
transition matrix elements (NTMEs) \cite{simk08,caur08,rath10a,rath11}.

Usually, three different approaches have been adopted for estimating the
uncertainties in NTMEs for $\left( \beta ^{-}\beta ^{-}\right) _{0\nu }$
decay. The spread between all the available calculated NTMEs has been used
as the measure of the theoretical uncertainty \cite{voge00}. The same spread
between NTMEs can also be translated into average and standard deviation,
which can be interpreted as theoretical uncertainty \cite{bahc04,avig05}.
According to Bilenky and Grifols \cite{bile02}, the observation of $\left(
\beta \beta \right) _{0\nu }$ decay of different nuclei will provide a
method, in which the ratios of the NTMEs-squared can be compared with the
ratios of observed half-lives $T_{1/2}^{0\nu }$ and the results of
calculations of NTMEs can be checked in a model independent way.

The theoretical uncertainties were estimated by Rodin \textit{et al.} \cite{rodi03} 
by considering two models, QRPA and RQRPA, with three sets of 
basis states and three realistic two-body effective interactions based on the charge 
dependent Bonn, Argonne and Nijmen potentials. It was found that the variances 
were substantially smaller than the average values and the results of QRPA, albeit 
slightly larger, are quite close to the RQRPA values. The critical analysis of the 
advantages and deficiencies in the approach of Rodin \textit{et al. }\cite{rodi03}
by Suhonen \cite{suho05} and Rodin $et$ $al.$ \cite{rodi06} is quite instructive. 
Further studies on the uncertainties in NTMEs due to SRC using the unitary correlation 
operator method (UCOM) \cite{kort07} and by self-consistent coupled cluster method (CCM) 
\cite{simk09} have also been carried out.

Recently, the uncertainties in the $\left( \beta ^{-}\beta ^{-}\right) _{0\nu }$ NTMEs
due to the exchange of light \cite{rath10a} and heavy \cite{rath11}
Majorana neutrinos have been calculated in the PHFB model by employing four
different parameterizations of the pairing plus multipolar effective two body
interaction and three different parameterizations of Jastrow type of SRC. 
In the present work, we employ the same formalism for estimating uncertainties in 
NTMEs for $(\varepsilon \beta
^{+})_{0\nu }$ and $(\varepsilon \varepsilon )_{0\nu }$ modes of $^{96}$Ru, $%
^{102}$Pd, $^{106}$Cd, $^{124}$Xe, $^{130}$Ba and $^{156}$Dy isotopes for
the $0^{+}\rightarrow 0^{+}$ transition. The article
is organized as follows. A brief discussion of the theoretical formalism is presented in 
Sec. II. In Sec. III, we analyze the role of the different parameterizations of
the two body interaction, the finite size of nucleons and higher order currents (HOC).
The influence of the SRC in the radial evolution of the NTMEs is also presented. 
In the same Sec. III, we estimate the uncertainties, which are subsequently employed 
for extracting bounds on the effective mass of light neutrinos 
$\left\langle m_{\nu }\right\rangle $ and heavy neutrinos 
$\left\langle M_{N}\right\rangle $. In Sec. IV, the conclusions are presented.

\section{THEORETICAL FORMALISM}

In the Majorana neutrino mass mechanism, the half-lives $T_{1/2}^{0\nu }$ for the
$0^{+} \to 0^{+}$\ transition of $\left( \beta ^{+}\beta ^{+}\right) _{0\nu
}$ and $\left( \varepsilon \beta ^{+}\right) _{0\nu }$ modes are given
by \cite{doi93,simk99,verg02}
\begin{equation}
\left[ T_{1/2}^{0\nu }\left( \beta \right) \right] ^{-1} =
G_{01}\left( \beta\right)\left|\frac{
\left\langle m_{\nu }\right\rangle }{m_{e}}
M^{\left( 0\nu \right) }+\frac{m_{p}}{
\left\langle M_{N}\right\rangle }
M^{\left( 0N\right) }\right|^2 .
\end{equation}
Here, $\beta $ denotes the $\left( \beta ^{+}\beta ^{+}\right) _{0\nu
}/\left( \varepsilon \beta ^{+}\right) _{0\nu }$ modes,
\begin{eqnarray}
\left\langle m_{\nu }\right\rangle &=&\sum\nolimits_{i}^{\prime
}U_{ei}^{2}m_{i},\qquad \qquad m_{i}<10\text{ }eV, \\
\left\langle M_{N}\right\rangle ^{-1} &=&\sum\nolimits_{i}^{\prime \prime
}U_{ei}^{2}m_{i}^{-1},\quad \qquad m_{i}>1\text{ }GeV.
\end{eqnarray}

\noindent 
and
\begin{equation}
M^{\left( K\right) }=- \frac{M_{F}^{\left( K\right)}}{ g_{A}^{2} } + M_{GT}^{\left( K\right)
}+M_{T}^{\left( K\right) }
\end{equation}
where $K = 0\nu \,(0N)$ denotes the exchange of light (heavy) Majorana neutrino mechanism.

In the PHFB model, the NTMEs $M^{\left( K\right) }$ for the $\left( \beta
^{+}\beta ^{+}\right) _{0\nu }$ and $\left( \varepsilon \beta ^{+}\right)
_{0\nu }$ modes are calculated by employing the closure approximation \cite
{rath09}
\begin{widetext}
\begin{eqnarray}
M^{\left( K\right) } &=&\langle \Psi {_{00}^{J_{f}=0}}||O^{\left(K\right) } 
||\Psi {_{00}^{J_{i}=0}}\rangle   \nonumber \\
&=&[n_{Z,N}^{J_{i}=0}n_{Z-2,N+2}^{J_{f}=0}]^{-1/2}\int\limits_{0}^{\pi
}n_{(Z,N),(Z-2,N+2)}(\theta )\sum\limits_{\alpha \beta \gamma \delta }\left(
\alpha \beta \left| O^{\left( K\right)}\right| \gamma \delta \right)   \nonumber \\
&&\times \sum_{\varepsilon \eta }\frac{(f_{Z-2,N+2}^{(\nu )*})_{\varepsilon
\beta }}{\left[ 1+F_{Z,N}^{(\nu )}(\theta )f_{Z-2,N+2}^{(\nu )*}\right]
_{\varepsilon \alpha }}\frac{(F_{Z,N}^{(\pi )*})_{\eta \delta }}{\left[
1+F_{Z,N}^{(\pi )}(\theta )f_{Z-2,N+2}^{(\pi )*}\right] _{\gamma \eta }}\sin
\theta d\theta \label{Mn}
\end{eqnarray}
where
\begin{eqnarray}
O^{(K)}=\left[ - \frac{H_{F}^{(K)}(r_{nm})}{g_{A}^{2}}+\mathbf{\sigma }_{n}\cdot
\mathbf{\sigma }_{m}H_{GT}^{(K)}(r_{nm})+S_{12}H_{T}^{(K)}(r_{nm})\right]
\tau _{n}^{+}\tau _{m}^{+} \label{TO}
\end{eqnarray}
with
\begin{eqnarray}
S_{nm}=3\left( \mathbf{\sigma }_{n}\cdot \widehat{\mathbf{r}}_{nm}\right)
\left( \mathbf{\sigma }_{m}\cdot \widehat{\mathbf{r}}_{nm}\right) -\mathbf{%
\sigma }_{n}\cdot \mathbf{\sigma }_{m}
\end{eqnarray}
\end{widetext}
and the expressions for $n^{J}$, $n_{(Z,N),(Z-2,N+2)}(\theta )$, $f_{Z,N}$\
\ and $F_{Z,N}(\theta )\ $are given in Ref. \cite{rath09}. The three components of 
the nuclear transition matrix element $M^{\left( K\right) }$ are denoted by 
$F$, $GT$ and $T$ corresponding to Fermi, Gamow-Teller and tensor terms.

The neutrino potentials due to the exchange of light and heavy neutrinos
between nucleons having finite size are given by
\begin{eqnarray}
H_{\alpha }^{\left( 0\nu \right) }(r_{nm}) &=&\frac{2R}{\pi }\int \frac{%
f_{\alpha }\left( qr_{nm}\right) }{\left( q+\overline{A}\right) }\;h_{\alpha
}(q)qdq \\  
H_{\alpha }^{\left( 0N\right) }(r_{nm}) &=&\frac{2R}{(m_{p}m_{e})\pi }\int
f_{\alpha }\left( qr_{nm}\right) h_{\alpha }(q)q^{2}dq
\end{eqnarray}
where $f_{\alpha }\left( qr_{nm}\right) =j_{0}\left( qr_{nm}\right) $  for $%
\alpha =F, GT$ and $f_{\alpha }\left( qr_{nm}\right) =j_{2}\left(
qr_{nm}\right) $ for $\alpha =T$. The above expressions for the NTMEs  
$M^{\left( K\right) }$ were obtained by including pseudoscalar and weak magnetism terms 
in the nucleonic current and employing the Goldberger-Treiman PCAC relation for the 
induced pseudoscalar term \cite{simk99}.

Usually, the influence of the finite size of nucleons (FNS) is taken into account through 
dipole form factors. The functions 
$h_{F}(q)$, $h_{GT}(q)$ 
and $h_{T}(q)$ are written as
\begin{widetext}
\begin{eqnarray}
h_{F}(q) &=&g_{V}^{2}(q^{2}) \\
h_{GT}(q) &=&\frac{g_{A}^{2}(q^{2})}{g_{A}^{2}}\left[ 1-\frac{2}{3}\frac{%
g_{P}(q^{2})q^{2}}{g_{A}(q^{2})2m_{p}}+\frac{1}{3}\frac{g_{P}^{2}(q^{2})q^{4}%
}{g_{A}^{2}(q^{2})4m_{p}^{2}}\right] +\frac{2}{3}\frac{g_{M}^{2}(q^{2})q^{2}%
}{g_{A}^{2}4m_{p}^{2}}  \nonumber \\
&\approx &\left( \frac{\Lambda _{A}^{2}}{q^{2}+\Lambda _{A}^{2}}\right)
^{4}\left[ 1-\frac{2}{3}\frac{q^{2}}{\left( q^{2}+m_{\pi }^{2}\right) }+%
\frac{1}{3}\frac{q^{4}}{\left( q^{2}+m_{\pi }^{2}\right) ^{2}}\right]
+\left( \frac{g_{V}}{g_{A}}\right) ^{2}\frac{\kappa ^{2}q^{2}}{6m_{p}^{2}}%
\left( \frac{\Lambda _{V}^{2}}{q^{2}+\Lambda _{V}^{2}}\right) ^{4} \\
h_{T}(q) &=&\frac{g_{A}^{2}(q^{2})}{g_{A}^{2}}\left[ \frac{2}{3}\frac{%
g_{P}(q^{2})q^{2}}{g_{A}(q^{2})2m_{p}}-\frac{1}{3}\frac{g_{P}^{2}(q^{2})q^{4}%
}{g_{A}^{2}(q^{2})4m_{p}^{2}}\right] +\frac{1}{3}\frac{g_{M}^{2}(q^{2})q^{2}%
}{g_{A}^{2}4m_{p}^{2}}  \nonumber \\
&\approx &\left( \frac{\Lambda _{A}^{2}}{q^{2}+\Lambda _{A}^{2}}\right)
^{4}\left[ \frac{2}{3}\frac{q^{2}}{\left( q^{2}+m_{\pi }^{2}\right) }-\frac{1%
}{3}\frac{q^{4}}{\left( q^{2}+m_{\pi }^{2}\right) ^{2}}\right] +\left( \frac{%
g_{V}}{g_{A}}\right) ^{2}\frac{\kappa ^{2}q^{2}}{12m_{p}^{2}}\left( \frac{%
\Lambda _{V}^{2}}{q^{2}+\Lambda _{V}^{2}}\right) ^{4} \label{hT}
\end{eqnarray}
\end{widetext}
where
\begin{eqnarray}
g_{V}(q^{2}) &=&g_{V}\left( \dfrac{\Lambda _{V}^{2}}{q^{2}+\Lambda _{V}^{2}}%
\right) ^{2}  \nonumber \\
g_{A}(q^{2}) &=&g_{A}\left( \dfrac{\Lambda _{A}^{2}}{q^{2}+\Lambda _{A}^{2}}%
\right) ^{2}  \nonumber \\
g_{M}(q^{2}) &=&\kappa g_{V}\left( q^{2}\right)   \nonumber \\
g_{P}(q^{2}) &=&\dfrac{2m_{p}g_{A}(q^{2})}{\left( q^{2}+m_{\pi }^{2}\right) }%
\left( \dfrac{\Lambda _{A}^{2}-m_{\pi }^{2}}{\Lambda _{A}^{2}}\right)
\end{eqnarray}
with $g_{V}=1.0$, $g_{A}=1.254$, $\kappa =\mu _{p}-\mu _{n}=3.70$, $\Lambda
_{V}=0.850$ GeV and $\Lambda _{A}=1.086$ GeV.

Consideration of Eq. (\ref{TO})--Eq. (\ref{hT}) and Eq. (\ref{Mn}) implies that the Fermi
matrix element $M_{F}^{ \left( K \right) }$ has one term 
-$g_{A}^{2}M_{F-VV}^{ \left( K \right) }$, the Gamow-Teller matrix element 
$M_{GT}^{ \left( K \right) }$ has four terms, namely $M_{GT-AA}^{ \left( K \right) }$, 
$M_{GT-AP}^{ \left( K \right) }$, $M_{GT-PP}^{ \left( K \right) }$, 
$M_{GT-MM}^{ \left( K \right) }$ and there are three terms $M_{T-AP}^{ \left( K \right) }$, 
$M_{T-PP}^{ \left( K \right) }$, $M_{T-MM}^{ \left( K \right) }$ associated 
with the tensor matrix element $M_{T}^{ \left( K \right) }$.

In the literature, the short range correlations (SRC) have been
included through the exchange of $\omega $-meson \cite{jghi95}, effective
transition operator \cite{wu85}, unitary correlation operator method (UCOM)
\cite{kort07,simk08}, self-consistent CCM \cite{simk09} and phenomenological
Jastrow type of correlations with Miller-Spenser parameterization \cite
{mill76}. Further, \v{S}imkovic \textit{et al.} \cite{simk09} have shown that in
the self-consistent CCM, it is possible to parametrize the effects of
Argonne V18 and CD-Bonn nucleon-nucleon $(NN)$ potentials by the Jastrow correlations 
with Miller-Spenser type of parameterization given by
\begin{equation}
f(r)=1-ce^{-ar^{2}}(1-br^{2}) .
\end{equation}
In the present work, the above form is adopted 
with $a=1.1$ $fm^{-2}$, $1.59$ $fm^{-2}$, $1.52$ $fm^{-2}$, $b=0.68$ $%
fm^{-2} $, $1.45$ $fm^{-2}$, $1.88$ $fm^{-2}$ and $c=1.0$, $0.92$, $0.46$
for Miller-Spencer parameterization, Argonne V18 and CD-Bonn $NN$ Potentials,
which are denoted as SRC1, SRC2 and SRC3, respectively.

The NTMEs $M^{\left(K \right) }$ of the $\left(
\beta ^{+}\beta ^{+}\right) _{0\nu }/\left( \varepsilon \beta ^{+}\right)
_{0\nu }$ decay mode in the PHFB model have been already discussed in Ref.
\cite{rath09}. The same formalism is employed here.
The axially symmetric HFB intrinsic state ${|\Phi _{0}\rangle }$ with $K=0$
specified completely by the amplitudes $(u_{im},v_{im})$ and expansion
coefficients $C_{ij,m}$, is obtained by minimizing the expectation value of
the effective Hamiltonian given by \cite{chan09}
\begin{equation}
H=H_{sp}+V(P)+V(QQ)+V(HH),
\end{equation}
in a basis constructed by using a set of deformed states. Here, $H_{sp}$ denotes 
the single particle Hamiltonian and $V(P)$, $V(QQ)$ and $V(HH)$ are the
pairing, quadrupole-quadrupole and hexadecapole-hexadecapole parts of the
effective two-body interaction, respectively.

The details about the parameters of the pairing force $G_{pp}$ and $G_{nn}$ as
well as three strength parameters of quadrupolar interaction, namely the
proton-proton $\chi _{2pp}$, the neutron-neutron $\chi _{2nn}$ and the
proton-neutron $\chi _{2pn}$ have been given in Refs. \cite
{rain06,sing07,rath09}. Specifically, $\chi _{2pp}=\chi _{2nn}=0.0105$ MeV$%
b^{-4}$, where $b$ is the oscillator parameter and the strength parameter $%
\chi _{2pn}$ was varied to fit the experimental excitation energy of the $\ $%
2$^{+}$ state, \ $E_{2^{+}}$. Presently, we employ in addition an
alternative isoscalar parameterization by taking $\chi _{2pp}=\chi
_{2nn}=\chi _{2pn}/2$ and the three parameters are varied together to fit $%
E_{2^{+}}$. These two parameterizations of the quadrupolar interaction are
referred as $PQQ1$ and $PQQ2$. The details about the $HH$ part of the
effective interaction $V(HH)$ have also been given in Ref. \cite{chan09}.
The calculations including the hexadecapolar term $HH$ are denoted as $PQQHH$%
. With the consideration of the hexadecapolar interaction, we end up with
four different parameterizations, namely $PQQ1$, $PQQHH1$, $PQQ2$ and $PQQHH2$
of the effective two-body interaction. By employing the four
different parameterization of the two body effective interaction and three
different parameterizations of SRC, sets of twelve NTMEs $M^{\left( 0\nu
\right) }$ and $M^{\left( 0N\right) }$ for the $\left( \beta ^{+}\beta
^{+}\right) _{0\nu }$ and $\left( \varepsilon \beta ^{+}\right) _{0\nu }$
modes are obtained using Eq. (\ref{Mn}) and subsequently, the mean and
standard deviations are calculated for estimating uncertainties
associated in the results of the present work.

\begin{table*}
\caption{Calculated NTMEs $M^{(0\nu )}$ and $M^{(0N )}$ in the PHFB
model with four different parameterization of effective two-body interaction, 
namely (a) $PQQ1$, (b) $PQQHH1$, (c) $PQQ2$ and (d) $PQQHH2$
and three different parameterizations of Jastrow type of SRC for the $\left(
\beta ^{+}\beta ^{+}\right)_{0\nu }$ and $\left(\varepsilon \beta ^{+}\right) _{0\nu }$
modes of $^{96}$Ru, $^{102}$Pd, $^{106}$Cd, $^{124}$Xe, $^{130}$Ba and $^{156}$Dy
isotopes due to the exchange of light as well as heavy Majorana neutrinos. See footnote 
on p.3 of Ref. \cite{rath10a} for further details.}
\label{tab1}
\begin{tabular}{lrcccccccccccccclcccccccc}
\hline\hline
Nuclei&~~~~~~  &~~~  & \multicolumn{13}{c}{Light neutrino exchange} &  &  &
\multicolumn{7}{c}{Heavy neutrino exchange} \\ \cline{4-16}\cline{19-25}
\multicolumn{1}{r}{} &  &  & F &  & \multicolumn{5}{c}{F+S} & ~~~~ &
\multicolumn{5}{c}{F+S($\overline{A}/2$)} & \multicolumn{1}{c}{} & ~~~~~ & F &  &
\multicolumn{5}{c}{F+S} \\ \cline{6-10}\cline{12-16}\cline{21-25}
\multicolumn{1}{r}{} &  &  &  &  & SRC1 &  & SRC2 &  & SRC3 &  & SRC1 &  &
SRC2 &  & SRC3 & \multicolumn{1}{c}{} &  &  &  & SRC1 &  & SRC2 &  & SRC3 \\
\hline
\multicolumn{1}{r}{$^{96}$Ru} & (a) &  & \multicolumn{1}{r}{4.7979} &  &
\multicolumn{1}{r}{4.1538} &  & \multicolumn{1}{r}{4.7331} &  &
\multicolumn{1}{r}{4.9191} & \multicolumn{1}{r}{} & \multicolumn{1}{r}{4.5739
} &  & \multicolumn{1}{r}{5.1844} &  & \multicolumn{1}{r}{5.3777} &
\multicolumn{1}{r}{} &  & \multicolumn{1}{r}{251.3551} &  &
\multicolumn{1}{r}{87.5090} &  & \multicolumn{1}{r}{150.6483} &  &
\multicolumn{1}{r}{205.2102} \\
\multicolumn{1}{r}{} & (b) &  & \multicolumn{1}{r}{4.7820} &  &
\multicolumn{1}{r}{4.1352} &  & \multicolumn{1}{r}{4.7164} &  &
\multicolumn{1}{r}{4.9032} & \multicolumn{1}{r}{} & \multicolumn{1}{r}{4.5501
} &  & \multicolumn{1}{r}{5.1626} &  & \multicolumn{1}{r}{5.3567} &
\multicolumn{1}{r}{} &  & \multicolumn{1}{r}{252.6125} &  &
\multicolumn{1}{r}{88.1241} &  & \multicolumn{1}{r}{151.5084} &  &
\multicolumn{1}{r}{206.2858} \\
\multicolumn{1}{r}{} & (c) &  & \multicolumn{1}{r}{4.8334} &  &
\multicolumn{1}{r}{4.1861} &  & \multicolumn{1}{r}{4.7686} &  &
\multicolumn{1}{r}{4.9555} & \multicolumn{1}{r}{} & \multicolumn{1}{r}{4.6107
} &  & \multicolumn{1}{r}{5.2245} &  & \multicolumn{1}{r}{5.4187} &
\multicolumn{1}{r}{} &  & \multicolumn{1}{r}{252.8381} &  &
\multicolumn{1}{r}{88.1649} &  & \multicolumn{1}{r}{151.6459} &  &
\multicolumn{1}{r}{206.4813} \\
\multicolumn{1}{r}{} & (d) &  & \multicolumn{1}{r}{4.7399} &  &
\multicolumn{1}{r}{4.1000} &  & \multicolumn{1}{r}{4.6753} &  &
\multicolumn{1}{r}{4.8601} & \multicolumn{1}{r}{} & \multicolumn{1}{r}{4.5120
} &  & \multicolumn{1}{r}{5.1182} &  & \multicolumn{1}{r}{5.3103} &
\multicolumn{1}{r}{} &  & \multicolumn{1}{r}{250.2061} &  &
\multicolumn{1}{r}{87.4415} &  & \multicolumn{1}{r}{150.1862} &  &
\multicolumn{1}{r}{204.3879} \\
\multicolumn{1}{r}{} &  &  & \multicolumn{1}{r}{} &  & \multicolumn{1}{r}{}
&  & \multicolumn{1}{r}{} &  & \multicolumn{1}{r}{} & \multicolumn{1}{r}{} &
\multicolumn{1}{r}{} &  & \multicolumn{1}{r}{} &  & \multicolumn{1}{r}{} &
\multicolumn{1}{r}{} &  & \multicolumn{1}{r}{} &  & \multicolumn{1}{r}{} &
& \multicolumn{1}{r}{} &  & \multicolumn{1}{r}{} \\
\multicolumn{1}{r}{$^{102}$Pd} & (a) &  & \multicolumn{1}{r}{5.3695} &  &
\multicolumn{1}{r}{4.5877} &  & \multicolumn{1}{r}{5.2981} &  &
\multicolumn{1}{r}{5.5230} & \multicolumn{1}{r}{} & \multicolumn{1}{r}{5.0512
} &  & \multicolumn{1}{r}{5.8007} &  & \multicolumn{1}{r}{6.0346} &
\multicolumn{1}{r}{} &  & \multicolumn{1}{r}{296.4236} &  &
\multicolumn{1}{r}{97.6551} &  & \multicolumn{1}{r}{174.1164} &  &
\multicolumn{1}{r}{240.2760} \\
\multicolumn{1}{r}{} & (b) &  & \multicolumn{1}{r}{4.5052} &  &
\multicolumn{1}{r}{3.8203} &  & \multicolumn{1}{r}{4.4407} &  &
\multicolumn{1}{r}{4.6377} & \multicolumn{1}{r}{} & \multicolumn{1}{r}{4.1853
} &  & \multicolumn{1}{r}{4.8400} &  & \multicolumn{1}{r}{5.0449} &
\multicolumn{1}{r}{} &  & \multicolumn{1}{r}{261.2562} &  &
\multicolumn{1}{r}{87.2707} &  & \multicolumn{1}{r}{154.2305} &  &
\multicolumn{1}{r}{212.1479} \\
\multicolumn{1}{r}{} & (c) &  & \multicolumn{1}{r}{5.4006} &  &
\multicolumn{1}{r}{4.6167} &  & \multicolumn{1}{r}{5.3292} &  &
\multicolumn{1}{r}{5.5547} & \multicolumn{1}{r}{} & \multicolumn{1}{r}{5.0831
} &  & \multicolumn{1}{r}{5.8348} &  & \multicolumn{1}{r}{6.0694} &
\multicolumn{1}{r}{} &  & \multicolumn{1}{r}{297.6968} &  &
\multicolumn{1}{r}{98.3279} &  & \multicolumn{1}{r}{175.0482} &  &
\multicolumn{1}{r}{241.4044} \\
\multicolumn{1}{r}{} & (d) &  & \multicolumn{1}{r}{4.4595} &  &
\multicolumn{1}{r}{3.7823} &  & \multicolumn{1}{r}{4.3959} &  &
\multicolumn{1}{r}{4.5907} & \multicolumn{1}{r}{} & \multicolumn{1}{r}{4.1432
} &  & \multicolumn{1}{r}{4.7907} &  & \multicolumn{1}{r}{4.9933} &
\multicolumn{1}{r}{} &  & \multicolumn{1}{r}{258.6259} &  &
\multicolumn{1}{r}{86.5570} &  & \multicolumn{1}{r}{152.7985} &  &
\multicolumn{1}{r}{210.0766} \\
\multicolumn{1}{r}{} &  &  & \multicolumn{1}{r}{} &  & \multicolumn{1}{r}{}
&  & \multicolumn{1}{r}{} &  & \multicolumn{1}{r}{} & \multicolumn{1}{r}{} &
\multicolumn{1}{r}{} &  & \multicolumn{1}{r}{} &  & \multicolumn{1}{r}{} &
\multicolumn{1}{r}{} &  & \multicolumn{1}{r}{} &  & \multicolumn{1}{r}{} &
& \multicolumn{1}{r}{} &  & \multicolumn{1}{r}{} \\
\multicolumn{1}{r}{$^{106}$Cd} & (a) &  & \multicolumn{1}{r}{8.4560} &  &
\multicolumn{1}{r}{7.2607} &  & \multicolumn{1}{r}{8.3403} &  &
\multicolumn{1}{r}{8.6835} & \multicolumn{1}{r}{} & \multicolumn{1}{r}{8.0547
} &  & \multicolumn{1}{r}{9.1947} &  & \multicolumn{1}{r}{9.5519} &
\multicolumn{1}{r}{} &  & \multicolumn{1}{r}{452.6855} &  &
\multicolumn{1}{r}{149.5474} &  & \multicolumn{1}{r}{265.9923} &  &
\multicolumn{1}{r}{366.9093} \\
\multicolumn{1}{r}{} & (b) &  & \multicolumn{1}{r}{6.9410} &  &
\multicolumn{1}{r}{5.9037} &  & \multicolumn{1}{r}{6.8370} &  &
\multicolumn{1}{r}{7.1347} & \multicolumn{1}{r}{} & \multicolumn{1}{r}{6.5165
} &  & \multicolumn{1}{r}{7.5021} &  & \multicolumn{1}{r}{7.8119} &
\multicolumn{1}{r}{} &  & \multicolumn{1}{r}{394.7635} &  &
\multicolumn{1}{r}{132.0821} &  & \multicolumn{1}{r}{233.0127} &  &
\multicolumn{1}{r}{320.4765} \\
\multicolumn{1}{r}{} & (c) &  & \multicolumn{1}{r}{8.5399} &  &
\multicolumn{1}{r}{7.3370} &  & \multicolumn{1}{r}{8.4229} &  &
\multicolumn{1}{r}{8.7683} & \multicolumn{1}{r}{} & \multicolumn{1}{r}{8.1436
} &  & \multicolumn{1}{r}{9.2902} &  & \multicolumn{1}{r}{9.6497} &
\multicolumn{1}{r}{} &  & \multicolumn{1}{r}{455.6637} &  &
\multicolumn{1}{r}{150.6151} &  & \multicolumn{1}{r}{267.7800} &  &
\multicolumn{1}{r}{369.3362} \\
\multicolumn{1}{r}{} & (d) &  & \multicolumn{1}{r}{7.7425} &  &
\multicolumn{1}{r}{6.6175} &  & \multicolumn{1}{r}{7.6293} &  &
\multicolumn{1}{r}{7.9524} & \multicolumn{1}{r}{} & \multicolumn{1}{r}{7.3228
} &  & \multicolumn{1}{r}{8.3911} &  & \multicolumn{1}{r}{8.7274} &
\multicolumn{1}{r}{} &  & \multicolumn{1}{r}{428.1103} &  &
\multicolumn{1}{r}{143.0153} &  & \multicolumn{1}{r}{252.5179} &  &
\multicolumn{1}{r}{347.4477} \\
\multicolumn{1}{r}{} &  &  & \multicolumn{1}{r}{} &  & \multicolumn{1}{r}{}
&  & \multicolumn{1}{r}{} &  & \multicolumn{1}{r}{} & \multicolumn{1}{r}{} &
\multicolumn{1}{r}{} &  & \multicolumn{1}{r}{} &  & \multicolumn{1}{r}{} &
\multicolumn{1}{r}{} &  & \multicolumn{1}{r}{} &  & \multicolumn{1}{r}{} &
& \multicolumn{1}{r}{} &  & \multicolumn{1}{r}{} \\
\multicolumn{1}{r}{$^{124}$Xe} & (a) &  & \multicolumn{1}{r}{4.1442} &  &
\multicolumn{1}{r}{3.5405} &  & \multicolumn{1}{r}{4.0770} &  &
\multicolumn{1}{r}{4.2507} & \multicolumn{1}{r}{} & \multicolumn{1}{r}{3.9471
} &  & \multicolumn{1}{r}{4.5153} &  & \multicolumn{1}{r}{4.6966} &
\multicolumn{1}{r}{} &  & \multicolumn{1}{r}{230.5375} &  &
\multicolumn{1}{r}{76.6774} &  & \multicolumn{1}{r}{135.6415} &  &
\multicolumn{1}{r}{186.8959} \\
\multicolumn{1}{r}{} & (b) &  & \multicolumn{1}{r}{3.4015} &  &
\multicolumn{1}{r}{2.8367} &  & \multicolumn{1}{r}{3.3342} &  &
\multicolumn{1}{r}{3.4963} & \multicolumn{1}{r}{} & \multicolumn{1}{r}{3.1370
} &  & \multicolumn{1}{r}{3.6639} &  & \multicolumn{1}{r}{3.8331} &
\multicolumn{1}{r}{} &  & \multicolumn{1}{r}{213.1788} &  &
\multicolumn{1}{r}{70.0146} &  & \multicolumn{1}{r}{124.7059} &  &
\multicolumn{1}{r}{172.4178} \\
\multicolumn{1}{r}{} & (c) &  & \multicolumn{1}{r}{3.6899} &  &
\multicolumn{1}{r}{3.1428} &  & \multicolumn{1}{r}{3.6275} &  &
\multicolumn{1}{r}{3.7849} & \multicolumn{1}{r}{} & \multicolumn{1}{r}{3.5024
} &  & \multicolumn{1}{r}{4.0157} &  & \multicolumn{1}{r}{4.1799} &
\multicolumn{1}{r}{} &  & \multicolumn{1}{r}{207.7699} &  &
\multicolumn{1}{r}{68.5459} &  & \multicolumn{1}{r}{121.8099} &  &
\multicolumn{1}{r}{168.1977} \\
\multicolumn{1}{r}{} & (d) &  & \multicolumn{1}{r}{3.4722} &  &
\multicolumn{1}{r}{2.8994} &  & \multicolumn{1}{r}{3.4045} &  &
\multicolumn{1}{r}{3.5690} & \multicolumn{1}{r}{} & \multicolumn{1}{r}{3.2056
} &  & \multicolumn{1}{r}{3.7406} &  & \multicolumn{1}{r}{3.9123} &
\multicolumn{1}{r}{} &  & \multicolumn{1}{r}{216.5439} &  &
\multicolumn{1}{r}{71.2281} &  & \multicolumn{1}{r}{126.7645} &  &
\multicolumn{1}{r}{175.1900} \\
\multicolumn{1}{r}{} &  &  & \multicolumn{1}{r}{} &  & \multicolumn{1}{r}{}
&  & \multicolumn{1}{r}{} &  & \multicolumn{1}{r}{} & \multicolumn{1}{r}{} &
\multicolumn{1}{r}{} &  & \multicolumn{1}{r}{} &  & \multicolumn{1}{r}{} &
\multicolumn{1}{r}{} &  & \multicolumn{1}{r}{} &  & \multicolumn{1}{r}{} &
& \multicolumn{1}{r}{} &  & \multicolumn{1}{r}{} \\
\multicolumn{1}{r}{$^{130}$Ba} & (a) &  & \multicolumn{1}{r}{3.5986} &  &
\multicolumn{1}{r}{3.0605} &  & \multicolumn{1}{r}{3.5369} &  &
\multicolumn{1}{r}{3.6914} & \multicolumn{1}{r}{} & \multicolumn{1}{r}{3.4204
} &  & \multicolumn{1}{r}{3.9254} &  & \multicolumn{1}{r}{4.0868} &
\multicolumn{1}{r}{} &  & \multicolumn{1}{r}{205.5885} &  &
\multicolumn{1}{r}{68.6378} &  & \multicolumn{1}{r}{121.1524} &  &
\multicolumn{1}{r}{166.7736} \\
\multicolumn{1}{r}{} & (b) &  & \multicolumn{1}{r}{2.8901} &  &
\multicolumn{1}{r}{2.4039} &  & \multicolumn{1}{r}{2.8372} &  &
\multicolumn{1}{r}{2.9769} & \multicolumn{1}{r}{} & \multicolumn{1}{r}{2.6473
} &  & \multicolumn{1}{r}{3.1067} &  & \multicolumn{1}{r}{3.2525} &
\multicolumn{1}{r}{} &  & \multicolumn{1}{r}{186.0534} &  &
\multicolumn{1}{r}{62.0004} &  & \multicolumn{1}{r}{109.6716} &  &
\multicolumn{1}{r}{150.9816} \\
\multicolumn{1}{r}{} & (c) &  & \multicolumn{1}{r}{2.9496} &  &
\multicolumn{1}{r}{2.4910} &  & \multicolumn{1}{r}{2.8950} &  &
\multicolumn{1}{r}{3.0266} & \multicolumn{1}{r}{} & \multicolumn{1}{r}{2.7799
} &  & \multicolumn{1}{r}{3.2083} &  & \multicolumn{1}{r}{3.3457} &
\multicolumn{1}{r}{} &  & \multicolumn{1}{r}{174.2028} &  &
\multicolumn{1}{r}{57.7641} &  & \multicolumn{1}{r}{102.3282} &  &
\multicolumn{1}{r}{141.1255} \\
\multicolumn{1}{r}{} & (d) &  & \multicolumn{1}{r}{1.5194} &  &
\multicolumn{1}{r}{1.2183} &  & \multicolumn{1}{r}{1.4801} &  &
\multicolumn{1}{r}{1.5662} & \multicolumn{1}{r}{} & \multicolumn{1}{r}{1.3358
} &  & \multicolumn{1}{r}{1.6133} &  & \multicolumn{1}{r}{1.7032} &
\multicolumn{1}{r}{} &  & \multicolumn{1}{r}{110.7898} &  &
\multicolumn{1}{r}{34.9642} &  & \multicolumn{1}{r}{63.7693} &  &
\multicolumn{1}{r}{89.0510} \\
\multicolumn{1}{r}{} &  &  & \multicolumn{1}{r}{} &  & \multicolumn{1}{r}{}
&  & \multicolumn{1}{r}{} &  & \multicolumn{1}{r}{} & \multicolumn{1}{r}{} &
\multicolumn{1}{r}{} &  & \multicolumn{1}{r}{} &  & \multicolumn{1}{r}{} &
\multicolumn{1}{r}{} &  & \multicolumn{1}{r}{} &  & \multicolumn{1}{r}{} &
& \multicolumn{1}{r}{} &  & \multicolumn{1}{r}{} \\
\multicolumn{1}{r}{$^{156}$Dy} & (a) &  & \multicolumn{1}{r}{2.1901} &  &
\multicolumn{1}{r}{1.9136} &  & \multicolumn{1}{r}{2.1712} &  &
\multicolumn{1}{r}{2.2513} & \multicolumn{1}{r}{} & \multicolumn{1}{r}{2.1304
} &  & \multicolumn{1}{r}{2.4044} &  & \multicolumn{1}{r}{2.4882} &
\multicolumn{1}{r}{} &  & \multicolumn{1}{r}{112.6022} &  &
\multicolumn{1}{r}{39.8413} &  & \multicolumn{1}{r}{68.3273} &  &
\multicolumn{1}{r}{92.5002} \\
\multicolumn{1}{r}{} & (b) &  & \multicolumn{1}{r}{1.9284} &  &
\multicolumn{1}{r}{1.6658} &  & \multicolumn{1}{r}{1.9097} &  &
\multicolumn{1}{r}{1.9857} & \multicolumn{1}{r}{} & \multicolumn{1}{r}{1.8415
} &  & \multicolumn{1}{r}{2.1010} &  & \multicolumn{1}{r}{2.1805} &
\multicolumn{1}{r}{} &  & \multicolumn{1}{r}{106.1906} &  &
\multicolumn{1}{r}{37.2359} &  & \multicolumn{1}{r}{64.1897} &  &
\multicolumn{1}{r}{87.1017} \\
\multicolumn{1}{r}{} & (c) &  & \multicolumn{1}{r}{2.5208} &  &
\multicolumn{1}{r}{2.1991} &  & \multicolumn{1}{r}{2.4982} &  &
\multicolumn{1}{r}{2.5913} & \multicolumn{1}{r}{} & \multicolumn{1}{r}{2.4520
} &  & \multicolumn{1}{r}{2.7702} &  & \multicolumn{1}{r}{2.8675} &
\multicolumn{1}{r}{} &  & \multicolumn{1}{r}{130.9372} &  &
\multicolumn{1}{r}{46.4436} &  & \multicolumn{1}{r}{79.5200} &  &
\multicolumn{1}{r}{107.5918} \\
\multicolumn{1}{r}{} & (d) &  & \multicolumn{1}{r}{2.5789} &  &
\multicolumn{1}{r}{2.2241} &  & \multicolumn{1}{r}{2.5526} &  &
\multicolumn{1}{r}{2.6551} & \multicolumn{1}{r}{} & \multicolumn{1}{r}{2.4674
} &  & \multicolumn{1}{r}{2.8168} &  & \multicolumn{1}{r}{2.9241} &
\multicolumn{1}{r}{} &  & \multicolumn{1}{r}{142.7229} &  &
\multicolumn{1}{r}{49.7900} &  & \multicolumn{1}{r}{86.0621} &  &
\multicolumn{1}{r}{116.9468} \\ \hline\hline
\end{tabular}
\end{table*}

\begin{table*}
\caption{Decomposition of NTMEs $M^{(0\nu )}$ and $M^{(0N )}$ for the 
 $\left(\beta ^{+}\beta ^{+}\right)_{0\nu }$ and 
$\left(\varepsilon \beta ^{+}\right) _{0\nu }$ modes of $^{106}$Cd including 
finite size effect (F) and SRC (F+S) for the $PQQ1$ parameterization.}
\label{tab2}
\begin{tabular}{cccccccccrrrccccc}
\hline\hline
NTMEs~~~~~~~~~~ & \multicolumn{7}{c}{Light neutrino exchange ($K=0\nu $)} &  & 
~~~~~~~ &
\multicolumn{7}{c}{Heavy neutrino exchange ($K=0N$)} \\
\cline{2-8}\cline{11-17}
\multicolumn{1}{l}{} & F &~~~~~  & \multicolumn{5}{c}{F+S} & \multicolumn{1}{r}{}
&  & F & ~~~~~ & \multicolumn{5}{c}{F+S} \\ \cline{4-8}\cline{13-17}
\multicolumn{1}{l}{} & \multicolumn{1}{r}{} & \multicolumn{1}{r}{} & SRC1 &
~~~~~& SRC2 &~~~~~  & SRC3 & \multicolumn{1}{r}{} &  &  &  & SRC1 &~~~~~  
& SRC2 &~~~~~  & SRC3
\\ \hline
\multicolumn{1}{l}{$M_{F}^{(K)}$} & \multicolumn{1}{r}{2.6164} &
\multicolumn{1}{r}{} & \multicolumn{1}{r}{2.2956} & \multicolumn{1}{r}{} &
\multicolumn{1}{r}{2.6164} & \multicolumn{1}{r}{} & \multicolumn{1}{r}{2.7064
} & \multicolumn{1}{r}{} &  & 133.9749 &  & \multicolumn{1}{r}{69.7853} &
\multicolumn{1}{r}{} & \multicolumn{1}{r}{105.3773} & \multicolumn{1}{r}{} &
\multicolumn{1}{r}{125.4157} \\
\multicolumn{1}{l}{$M_{GT-AA}^{(K)}$} & \multicolumn{1}{r}{-7.8995} &
\multicolumn{1}{r}{} & \multicolumn{1}{r}{-6.7491} & \multicolumn{1}{r}{} &
\multicolumn{1}{r}{-7.7996} & \multicolumn{1}{r}{} & \multicolumn{1}{r}{
-8.1279} & \multicolumn{1}{r}{} &  & -461.3370 &  & \multicolumn{1}{r}{
-179.9510} & \multicolumn{1}{r}{} & \multicolumn{1}{r}{-301.7550} &
\multicolumn{1}{r}{} & \multicolumn{1}{r}{-393.8500} \\
\multicolumn{1}{l}{$M_{GT-AP}^{(K)}$} & \multicolumn{1}{r}{-.5562} &
\multicolumn{1}{r}{} & \multicolumn{1}{r}{-.3643} & \multicolumn{1}{r}{} &
\multicolumn{1}{r}{-.5003} & \multicolumn{1}{r}{} & \multicolumn{1}{r}{-.5570
} & \multicolumn{1}{r}{} &  & 218.6080 &  & \multicolumn{1}{r}{55.4365} &
\multicolumn{1}{r}{} & \multicolumn{1}{r}{118.4700} & \multicolumn{1}{r}{} &
\multicolumn{1}{r}{172.6570} \\
\multicolumn{1}{l}{$M_{GT-PP}^{(K)}$} & \multicolumn{1}{r}{1.8965} &
\multicolumn{1}{r}{} & \multicolumn{1}{r}{1.3980} & \multicolumn{1}{r}{} &
\multicolumn{1}{r}{1.7872} & \multicolumn{1}{r}{} & \multicolumn{1}{r}{1.9323
} & \multicolumn{1}{r}{} &  & -83.4189 &  & \multicolumn{1}{r}{-11.3250} &
\multicolumn{1}{r}{} & \multicolumn{1}{r}{-36.4994} & \multicolumn{1}{r}{} &
\multicolumn{1}{r}{-60.7425} \\
\multicolumn{1}{l}{$M_{GT-MM}^{(K)}$} & \multicolumn{1}{r}{-.2945} &
\multicolumn{1}{r}{} & \multicolumn{1}{r}{-.1479} & \multicolumn{1}{r}{} &
\multicolumn{1}{r}{-.2276} & \multicolumn{1}{r}{} & \multicolumn{1}{r}{-.2735
} & \multicolumn{1}{r}{} &  & -52.2544 &  & \multicolumn{1}{r}{20.1072} &
\multicolumn{1}{r}{} & \multicolumn{1}{r}{9.1939} & \multicolumn{1}{r}{} &
\multicolumn{1}{r}{-16.8909} \\
\multicolumn{1}{l}{$M_{GT}^{(K)}$} & \multicolumn{1}{r}{-6.8537} &
\multicolumn{1}{r}{} & \multicolumn{1}{r}{-5.8634} & \multicolumn{1}{r}{} &
\multicolumn{1}{r}{-6.7404} & \multicolumn{1}{r}{} & \multicolumn{1}{r}{
-7.0261} & \multicolumn{1}{r}{} &  & -378.4023 &  & \multicolumn{1}{r}{
-115.7323} & \multicolumn{1}{r}{} & \multicolumn{1}{r}{-210.5905} &
\multicolumn{1}{r}{} & \multicolumn{1}{r}{-298.8264} \\
\multicolumn{1}{l}{$M_{T-AP}^{(K)}$} & \multicolumn{1}{r}{-.0306} &
\multicolumn{1}{r}{} & \multicolumn{1}{r}{-.0311} & \multicolumn{1}{r}{} &
\multicolumn{1}{r}{-.0319} & \multicolumn{1}{r}{} & \multicolumn{1}{r}{-.0318
} & \multicolumn{1}{r}{} &  & 14.8474 &  & \multicolumn{1}{r}{14.5508} &
\multicolumn{1}{r}{} & \multicolumn{1}{r}{15.8270} & \multicolumn{1}{r}{} &
\multicolumn{1}{r}{15.8681} \\
\multicolumn{1}{l}{$M_{T-PP}^{(K)}$} & \multicolumn{1}{r}{.0835} &
\multicolumn{1}{r}{} & \multicolumn{1}{r}{.0849} & \multicolumn{1}{r}{} &
\multicolumn{1}{r}{.0866} & \multicolumn{1}{r}{} & \multicolumn{1}{r}{.0864}
& \multicolumn{1}{r}{} &  & -5.9893 &  & \multicolumn{1}{r}{-5.8221} &
\multicolumn{1}{r}{} & \multicolumn{1}{r}{-6.4169} & \multicolumn{1}{r}{} &
\multicolumn{1}{r}{-6.4410} \\
\multicolumn{1}{l}{$M_{T-MM}^{(K)}$} & \multicolumn{1}{r}{.0087} &
\multicolumn{1}{r}{} & \multicolumn{1}{r}{.0087} & \multicolumn{1}{r}{} &
\multicolumn{1}{r}{.0092} & \multicolumn{1}{r}{} & \multicolumn{1}{r}{.0091}
& \multicolumn{1}{r}{} &  & 2.0565 &  & \multicolumn{1}{r}{1.8336} &
\multicolumn{1}{r}{} & \multicolumn{1}{r}{2.2001} & \multicolumn{1}{r}{} &
\multicolumn{1}{r}{2.2449} \\
\multicolumn{1}{l}{$M_{T}^{(K)}$} & \multicolumn{1}{r}{.0615} &
\multicolumn{1}{r}{} & \multicolumn{1}{r}{.0625} & \multicolumn{1}{r}{} &
\multicolumn{1}{r}{.0638} & \multicolumn{1}{r}{} & \multicolumn{1}{r}{.0637}
& \multicolumn{1}{r}{} &  & 10.9146 &  & \multicolumn{1}{r}{10.5624} &
\multicolumn{1}{r}{} & \multicolumn{1}{r}{11.6101} & \multicolumn{1}{r}{} &
\multicolumn{1}{r}{11.6720} \\
\multicolumn{1}{l}{$\left| M^{(K)}\right| $} & \multicolumn{1}{r}{8.4560} &
\multicolumn{1}{r}{} & \multicolumn{1}{r}{7.2607} & \multicolumn{1}{r}{} &
\multicolumn{1}{r}{8.3403} & \multicolumn{1}{r}{} & \multicolumn{1}{r}{8.6835
} & \multicolumn{1}{r}{} &  & 452.6855 &  & \multicolumn{1}{r}{149.5474} &
\multicolumn{1}{r}{} & \multicolumn{1}{r}{265.9923} & \multicolumn{1}{r}{} &
\multicolumn{1}{r}{366.9093} \\ \hline\hline
\end{tabular}
\end{table*}

\section{RESULTS AND DISCUSSIONS}

In the present work, we use the same model space, single particle energies
(SPE's) and parameters of the effective two-body interaction as our
earlier calculations on the $\left( e^{+}\beta \beta \right) _{2\nu }$ decay
of $^{96}$Ru, $^{102}$Pd, $^{106,108}$Cd \cite{rain06}, $^{124,126}$Xe, $%
^{130,132}$Ba \cite{sing07} and $^{156}$Dy \cite{rath10b} isotopes for the $%
0^{+}\to 0^{+}$ transition.  
The calculated yrast
spectra, reduced $B(E2$:$0^{+}\to 2^{+})$ transition probabilities, static
quadrupole moments $Q(2^{+})$ and gyromagnetic factors $g(2^{+})$ \cite
{rain06,sing07,rath10b} are
in an overall agreement with the experimental data due to $PQQ1$ 
parameterization. The maximum change in all the calculated
spectroscopic properties is about 10\% except for the $B(E2$:$0^{+}\to 2^{+})
$ and $g(2^{+})$ of $^{130}$Ba, which change by 19.5\% and 16.6\%,
respectively, employing the other three parameterizations.

\subsection{Effects due to finite size of nucleons and short range correlations}

The theoretically calculated sets of twelve NTMEs $M^{\left( 0\nu \right) }$
and $M^{\left( 0N \right) }$ using the HFB\ wave functions in
conjunction with $PQQ1$, $PQQHH1$, $PQQ2$ and $PQQHH2$ interaction and three
different parameterizations of the Jastrow type of SRC for 
$^{96}$Ru, $^{102}$Pd, $^{106}$Cd, $^{124}$Xe, $^{130}$Ba 
and $^{156}$Dy nuclei are given in Table~\ref{tab1}. The sets of twelve NTMEs
$M^{\left( 0\nu \right) }$ and $M^{\left( 0N \right) }$ are calculated
in the approximation of finite size of nucleons with dipole form factor (F) and finite
size plus SRC (F+S). Further, the NTMEs $M^{\left( 0\nu \right) }$ are
calculated for $\overline{A}$ and $\overline{A}/2$ in the energy denominator
in the case of F+S. We present the components of NTMEs $M^{\left( 0\nu \right) }$ 
as well as $M^{\left( 0N \right) }$, namely Fermi, Gamow-Teller and tensor matrix 
elements of $^{106}$Cd in Table~\ref{tab2} for explicitly displaying the role of 
higher order currents (HOC). In Table~\ref{tab3}, the changes in NTMEs $%
M^{\left( 0\nu \right) }$ and $M^{\left( 0N \right) }$, due to different 
approximations are displayed. The following observations are noteworthy.

\begin{table*}[htbp]
\caption {Changes (in \%) of the 
NTMEs $M^{\left( 0\nu\right) }$ and $M^{\left( 0N\right) }$ 
due to exchange of light and heavy Majorana neutrinos, respectively, 
for the $\left( \beta ^{+}\beta ^{+}\right) _{0\nu }$ and $ \left(
\varepsilon \beta ^{+}\right) _{0\nu }$ modes with the inclusion of finite size 
effect (FNS) as well as finite size effect+HOC (F), F+SRC (F+SRC1, F+SRC2 and F+SRC3) 
for four different parameterizations of the effective two-body interaction.}
\label{tab3}
\begin{tabular}{lccccrcccccc}
\hline\hline
& \multicolumn{5}{c}{\small Light neutrino exchange} &  & \multicolumn{5}{c}%
{\small Heavy neutrino exchange} \\ \cline{2-6}\cline{8-12}
~~~~~~~~~~~~& {\small FNS} &~~~~~~~~  & {\small F} &~~~~~~~~  & {\small F+S} 
&~~~~~~~~~~~~  & {\small FNS} &~~~~~~~~  &
{\small F} &~~~~~~~~  & {\small F+S} \\ \hline
{\small (a)} & {\small 9.08--10.31} &  & {\small 10.91--12.47} &  & {\small %
12.63--14.95} &  & {\small 25.89--28.91} &  & {\small 14.41--17.51} &  &
{\small 64.62--67.06} \\
&  &  &  &  & {\small 0.86--1.72} &  &  &  &  &  & {\small 39.32--41.26} \\
&  &  &  &  & {\small 2.53--2.86} &  &  &  &  &  & {\small 17.85--18.95} \\
{\small (b)} & {\small 9.58--11.34} &  & {\small 10.87--13.39} &  & {\small %
13.53--16.82} &  & {\small 26.03--29.23} &  & {\small 14.25--16.75} &  &
{\small 64.93--67.16} \\
&  &  &  &  & {\small 0.97--1.98} &  &  &  &  &  & {\small 39.55--41.50} \\
&  &  &  &  & {\small 2.53--3.00} &  &  &  &  &  & {\small 17.98--19.12} \\
{\small (c)} & {\small 9.18--10.63} &  & {\small 10.88--12.43} &  & {\small %
12.76--15.55} &  & {\small 25.96--29.17} &  & {\small 14.35--17.40} &  &
{\small 64.53--67.01} \\
&  &  &  &  & {\small 0.89--1.85} &  &  &  &  &  & {\small 39.27--41.37} \\
&  &  &  &  & {\small 2.53--2.85} &  &  &  &  &  & {\small 17.83--19.05} \\
{\small (d)} & {\small 9.57--12.74} &  & {\small 10.84--14.91} &  & {\small %
13.50--19.81} &  & {\small 26.18--29.80} &  & {\small 14.19--17.46} &  &
{\small 65.05--68.44} \\
&  &  &  &  & {\small 1.02--2.59} &  &  &  &  &  & {\small 39.70--42.44} \\
&  &  &  &  & {\small 2.54--3.08} &  &  &  &  &  & {\small 18.06--19.62} \\
\hline\hline
\end{tabular}
\end{table*}

\begin{enumerate}[(i)]
\item  Changing $\overline{A}$ to  $\overline{A}/2$ in the energy denominator, 
the changes in the  NTMEs $M^{\left( 0\nu \right) }$ vary between 9\%--12 $\%$ 
exhibiting that the dependence of NTMEs on average excitation energy $\overline{A}$ is 
small, which supports the use of the closure approximation in the case of 
the $\left( \beta \beta \right) _{0\nu }$ decay.

\item 
Inclusion of effects due to FNS induces changes in the NTMEs $M_{VV}^{\left( 0\nu
\right) }+M_{AA}^{\left( 0\nu\right) }$ by 9.0\%--13.0\%. 
Further, the addition of higher order currents (HOC) reduces the NTMEs by 
11.0\%--15.0\%.

\item  With the addition of SRC1, SRC2 and SRC3, the NTMEs $M^{\left( 0\nu
\right) }$ vary approximately by 13\%--20\%, 0.9\%--2.6\% and 2.5\%--3.0\%,
respectively, in comparison to the case F.

\item The NTMEs $M_{VV}^{\left( 0N\right) }+M_{AA}^{\left( 0N\right) }$ in the case of 
heavy neutrino exchange, with the consideration of 
FNS instead of point nucleons, vary by 18.0\%--29.0\%, and the inclusion of  HOC results 
in further reduction by about 14.0\%--17.5\%. 

\item  In the case of heavy neutrino exchange, the NTMEs $M^{\left( 0N \right) }$ 
become smaller by approximately 65\%--68\%, 39\%--42\% and 18\%--20\% for SRC1, SRC2 and SRC3,
respectively. To understand the behaviour of SRC, we plot in Fig. 1 the neutrino potential
$H_{N}\left( r,\Lambda \right)$=$H^{\left (0N \right) }_{F}\left( r,\Lambda \right)f^{2}(r)$
with three different parameterizations of the SRC. The potential including only 
FNS is peaked at origin whereas the peaks due to F+SRC1, F+SRC2
and F+SRC3 are at $r\approx 0.8$ fm, 0.7 fm and 0.5 fm, respectively. The visible 
reduction of the area under the curves is the main cause behind the large changes 
reported in Table III.

\begin{figure}[htbp]
\includegraphics [scale=0.7,angle=270]{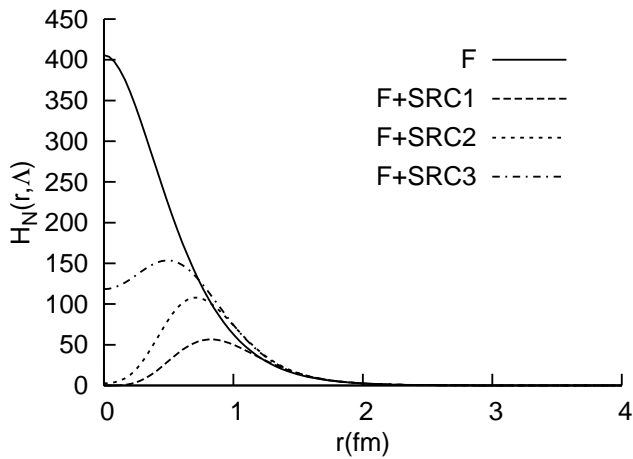}
\caption{Radial dependence of $H_{N}\left( r,\Lambda \right)$=
$H^{\left (0N \right) }_{F}\left( r,\Lambda \right)f^{2}(r)$ for the three different 
parameterizations of the SRC. In the case of FNS, $f(r)=1$.}
\label{fig1}
\end{figure}

\item  The maximum variations in $M^{\left( 0\nu \right) }$ ($M^{\left( 0N \right) }$)
due to $PQQHH1$, $PQQ2$ and $PQQHH2$ parameterizations with respect to $PQQ1$ interaction, 
but for the pathological case $^{130}$Ba, are about 21.0\% (13.0\%), 19.0\% (17.0\%) 
and 18.0\% (27.0\%). 

\item The effect of deformation on $M^{\left( K\right) }$ is 
quantified by the quantity $D^{\left( K\right) }$ as the ratio of $M^{\left( K\right) }$ 
at zero deformation ($\zeta _{qq}=0$) and full
deformation ($\zeta _{qq}=1$) and is given by \cite{rath09} 
\begin{equation}
D^{\left( K\right) }=\frac{M^{\left( K\right) }(\zeta
_{qq}=0)}{M^{\left( K\right) }(\zeta _{qq}=1)}
\end{equation}
In Table~\ref{tab4}, we tabulate the values of $D^{\left( K\right) }$ due
to exchange of light and heavy neutrinos for $^{96}$Ru, $^{102}$Pd, $^{106}$%
Cd, $^{124}$Xe, $^{130}$Ba and $^{156}$Dy nuclei. It is observed that, due to
deformation effects, the NTMEs $M^{\left( K\right) }$ are
suppressed by factor of 1.7--10.8 in the mass range $A=96-156$. Thus, 
the deformation plays a crucial role in the $\left( \beta ^{+}\beta
^{+}\right) _{0\nu }$ and $\left( \varepsilon \beta ^{+}\right) _{0\nu }$ modes.

\item It is also observed that excluding the pathological case $^{130}$Ba the 
ratios of NTMEs $M^{\left( 0\nu \right) }$/$M^{\left( 0N \right) }$ are 
about 21--25, 32--37 and 42--49 for SRC1, SRC2 and SRC3, respectively. The spread 
in the above mentioned ratios increases to 21--29, 32--43 and 42--57 with the 
consideration of $^{130}$Ba isotope.  

\end{enumerate}

\begin{table}[htbp]
\caption{Deformation ratios $\left( i\right) D^{\left( 0\nu \right) }$ and $%
\left( ii\right) D^{\left( 0N\right) }$ of  $\left( \beta ^{+}\beta
^{+}\right) _{0\nu }$ and $\left( \varepsilon \beta ^{+}\right) _{0\nu }$ modes 
for the $PQQ1$ parameterization.}
\label{tab4}
\begin{tabular}{lllllrrrr}
\hline\hline
Nuclei~~ &  &  &  &  & F & \multicolumn{3}{c}{ F+SRC} \\ \cline{7-8}\cline{9-9}
&  &  &  &  &  &~~ F+SRC1 &~~ F+SRC2 &~~ F+SRC3 \\ \hline
\multicolumn{1}{r}{$^{96}$Ru}~~ &  & $\left( i\right) $ &  &  & 2.53 & 2.54 &
2.53 & 2.53 \\
\multicolumn{1}{r}{} &  & $\left( ii\right) $ &  &  & 2.44 & 2.34 & 2.40 &
2.43 \\
\multicolumn{1}{r}{$^{102}$Pd}~~ &  & $\left( i\right) $ &  &  & 2.68 & 2.74 &
2.68 & 2.67 \\
\multicolumn{1}{r}{} &  & $\left( ii\right) $ &  &  & 2.40 & 2.45 & 2.42 &
2.40 \\
\multicolumn{1}{r}{$^{106}$Cd}~~ &  & $\left( i\right) $ &  &  & 1.92 & 1.96 &
1.93 & 1.92 \\
\multicolumn{1}{r}{} &  & $\left( ii\right) $ &  &  & 1.72 & 1.77 & 1.74 &
1.73 \\
\multicolumn{1}{r}{$^{124}$Xe}~~ &  & $\left( i\right) $ &  &  & 3.82 & 3.92 &
3.84 & 3.82 \\
\multicolumn{1}{r}{} &  & $\left( ii\right) $ &  &  & 3.36 & 3.50 & 3.42 &
3.38 \\
\multicolumn{1}{r}{$^{130}$Ba}~~ &  & $\left( i\right) $ &  &  & 4.61 & 4.75 &
4.65 & 4.61 \\
\multicolumn{1}{r}{} &  & $\left( ii\right) $ &  &  & 3.97 & 4.15 & 4.06 &
4.01 \\
\multicolumn{1}{r}{$^{156}$Dy}~~ &  & $\left( i\right) $ &  &  & 10.78 & 10.82
& 10.76 & 10.75 \\
\multicolumn{1}{r}{} &  & $\left( ii\right) $ &  &  & 10.25 & 10.01 & 10.14
& 10.20 \\ \hline\hline
\end{tabular}
\end{table}

\subsection{Radial evolution of NTMEs}

In the Majorana neutrino mass mechanism, the radial evolution of NTMEs can
be studied by defining 
\begin{equation}
M^{\left(K \right) }=\int C^{\left(K \right) }\left( r\right) dr.
\end{equation}
The study of radial evolution of NTMEs $M^{\left( 0\nu \right) }$ in the
QRPA by \v{S}imkovic $et$ $al.$ \cite{simk08}, and in the ISM by Men\'{e}ndez 
$et$ $al.$ \cite{mene09}, has established that the magnitude of $C^{\left(
0\nu \right) }$ for all nuclei undergoing $\left( \beta ^{-}\beta
^{-}\right) _{0\nu }$ decay exhibit a maximum at about the internucleon distance 
$r\approx 1$ fm, and that the contributions of decaying pairs coupled to $J=0$ 
and $J>0$ almost cancel out beyond $r\approx 3$ fm. In the PHFB model, the radial
evolution of NTMEs $M^{\left( 0\nu \right) }$ and $M^{\left( 0N
\right) }$ for $\left( \beta ^{-}\beta ^{-}\right) _{0\nu }$ decay due to
the exchange of light \cite{rath10a} and heavy Majorana neutrinos \cite{rath11}
has also been studied and similar observations have been reported.

\begin{figure}[htbp]
\begin{tabular}{c}
\includegraphics [scale=0.38]{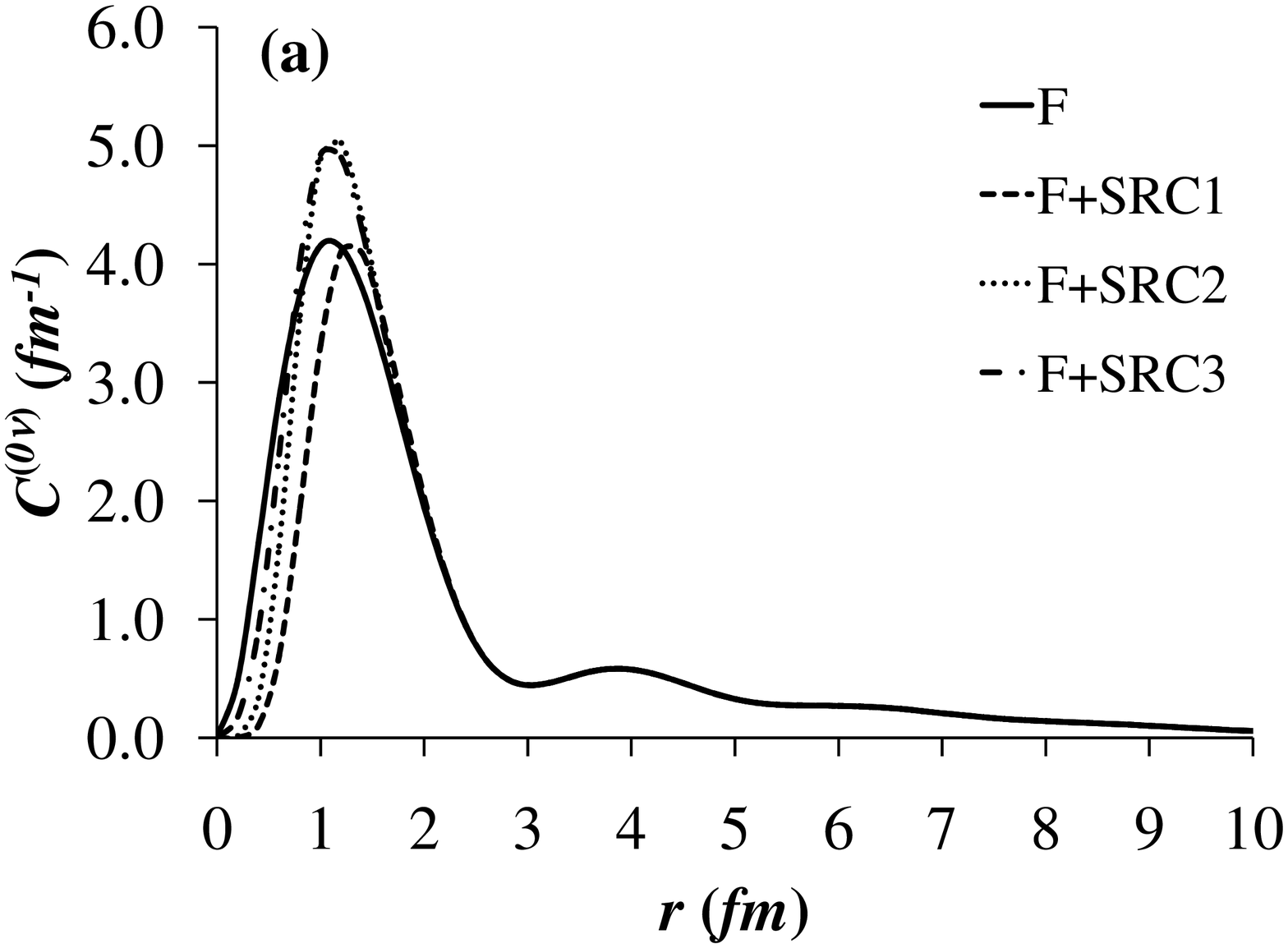} \\
\includegraphics [scale=0.38]{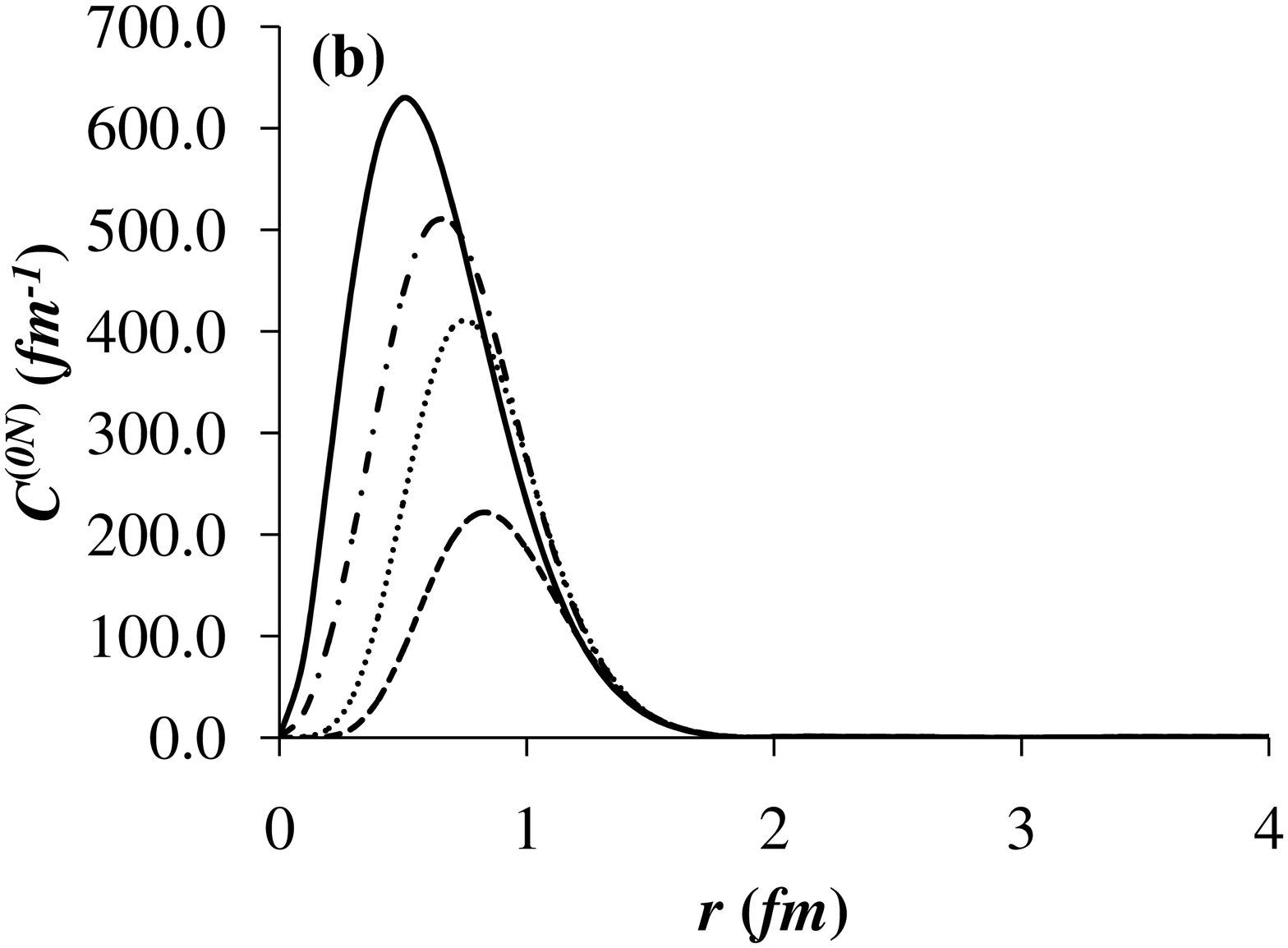} \\
\end{tabular}
\caption{Radial dependence of $C^{(0\nu)}(r)$ and $C^{(0N)}(r)$ for the
$\left( \beta ^{+}\beta ^{+} \right) _{0\nu }$ and
$\left( \varepsilon \beta ^{+}\right) _{0\nu }$ modes of $^{106}$Cd isotope.}
\label{fig2}
\end{figure}

\begin{figure*}[htbp]
\begin{tabular}{cc}
\includegraphics [scale=0.33]{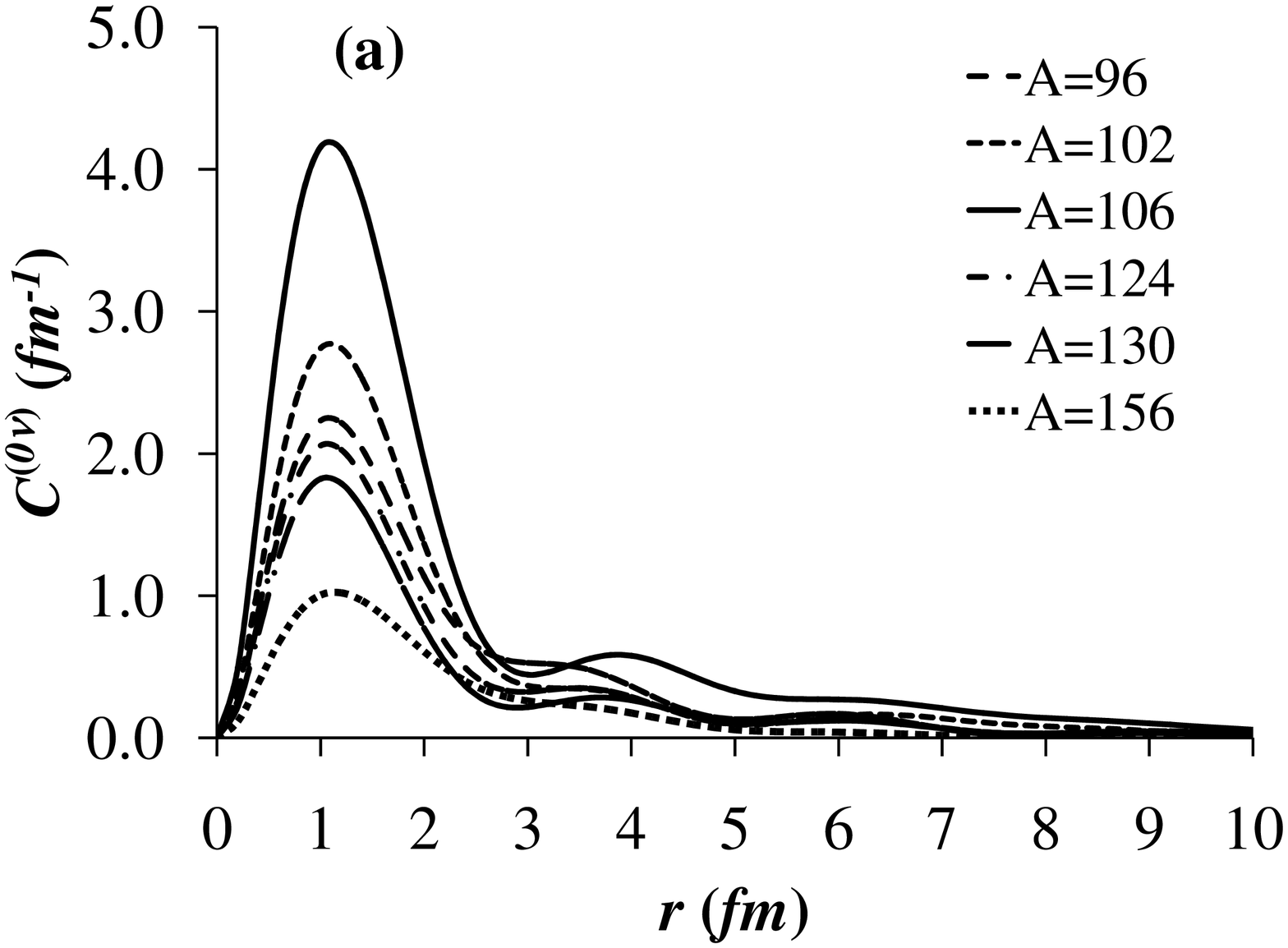} &
\includegraphics [scale=0.33]{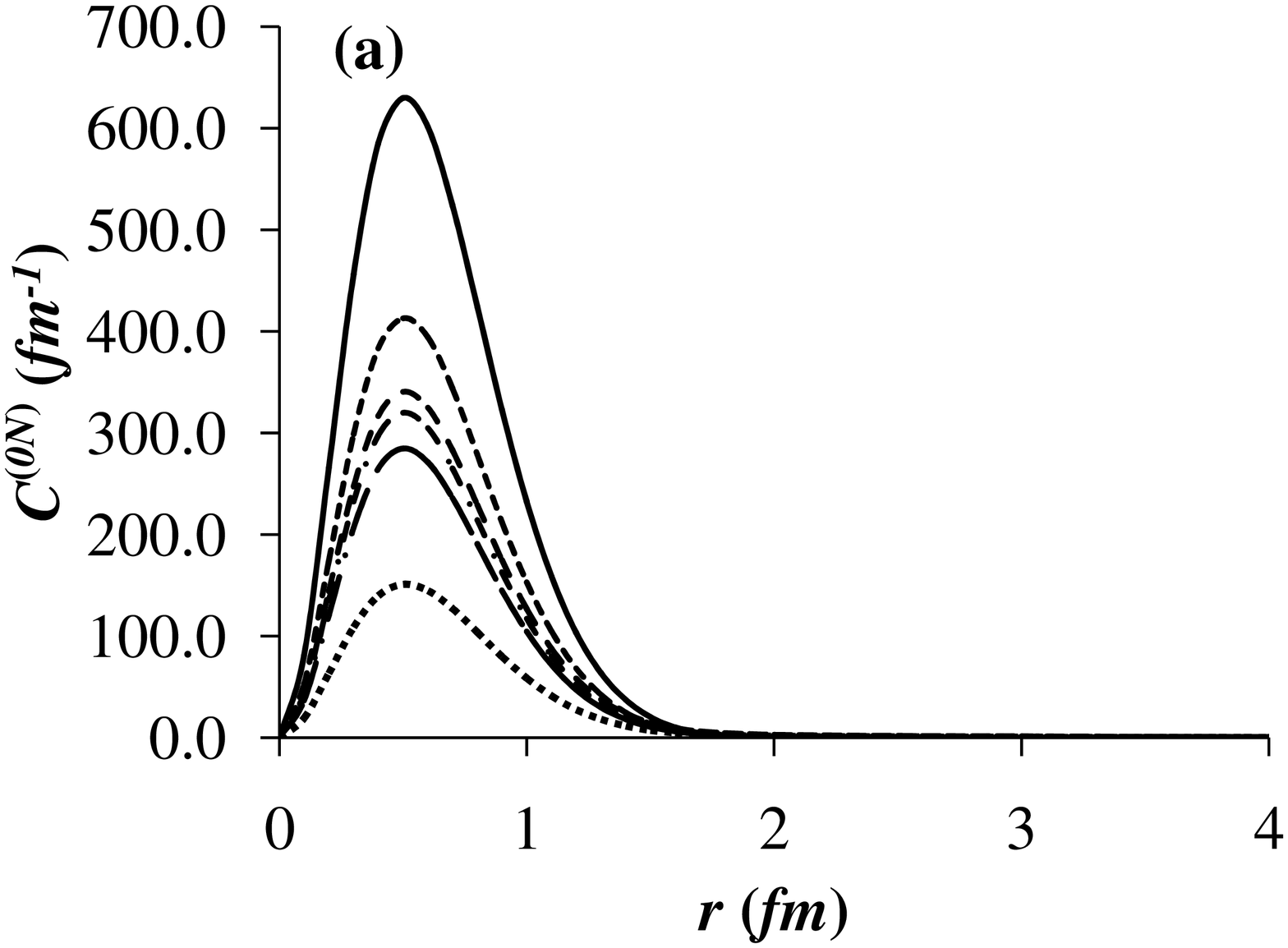} \\
\includegraphics [scale=0.33]{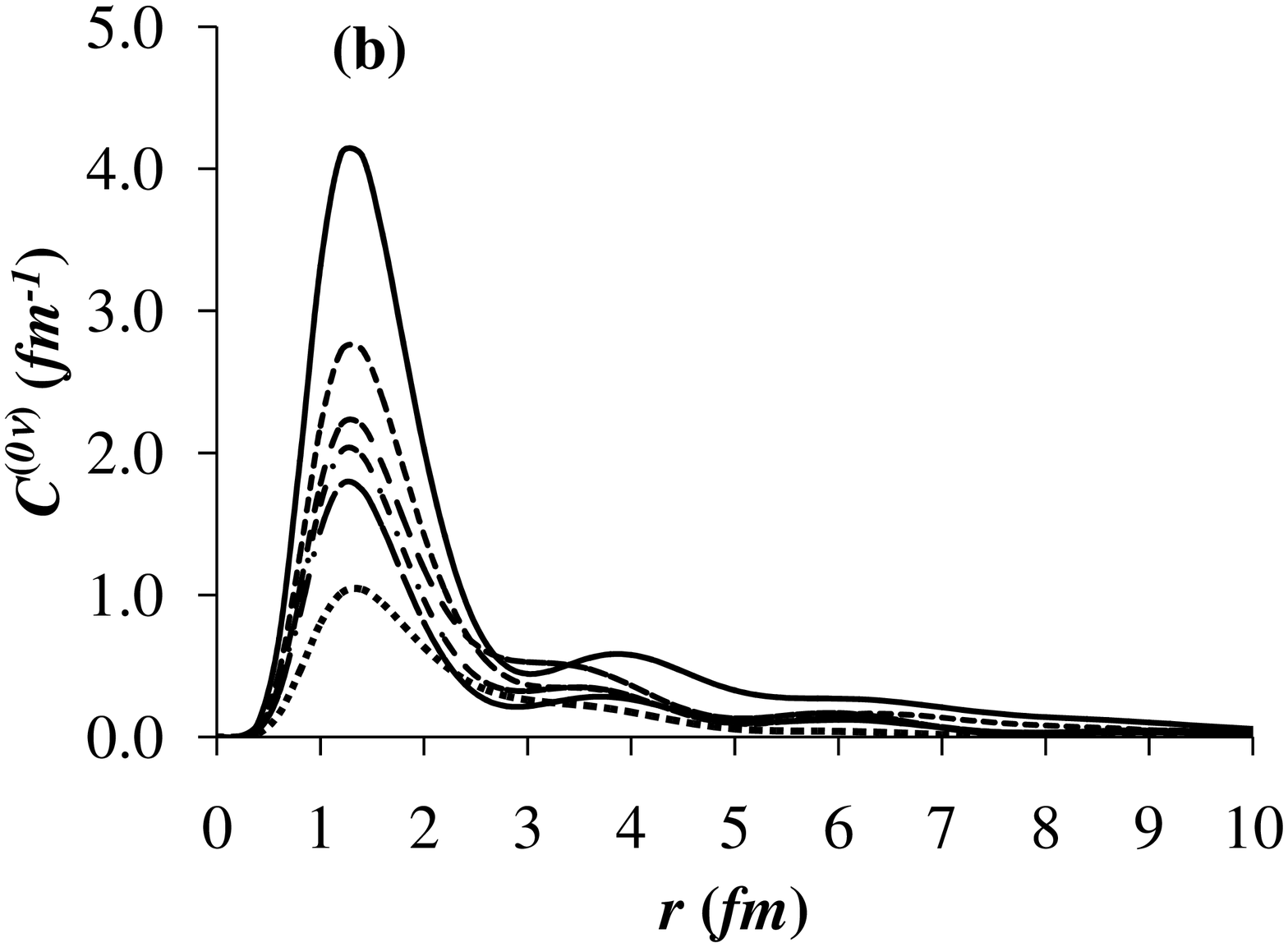} &
\includegraphics [scale=0.33]{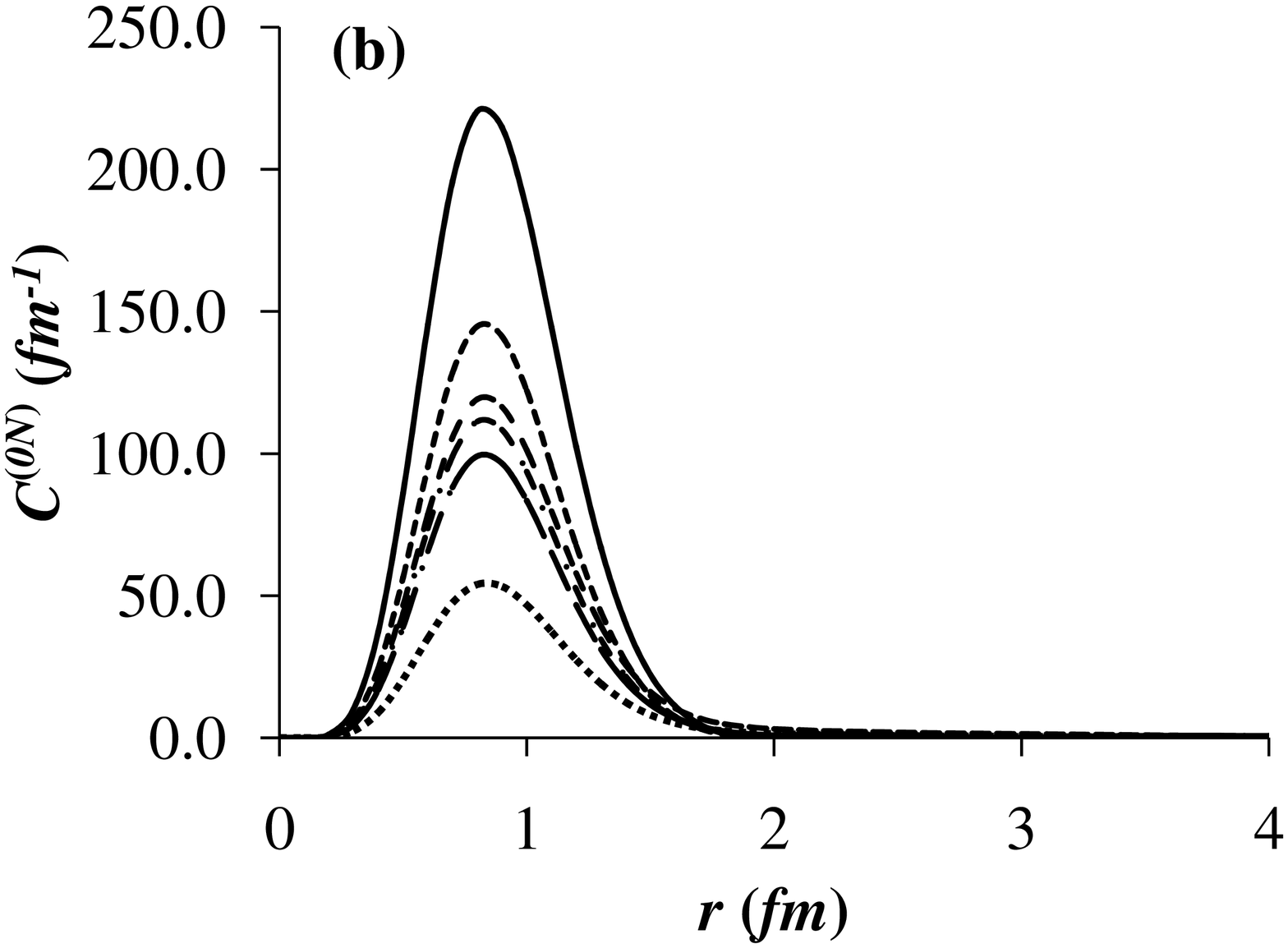} \\
\includegraphics [scale=0.33]{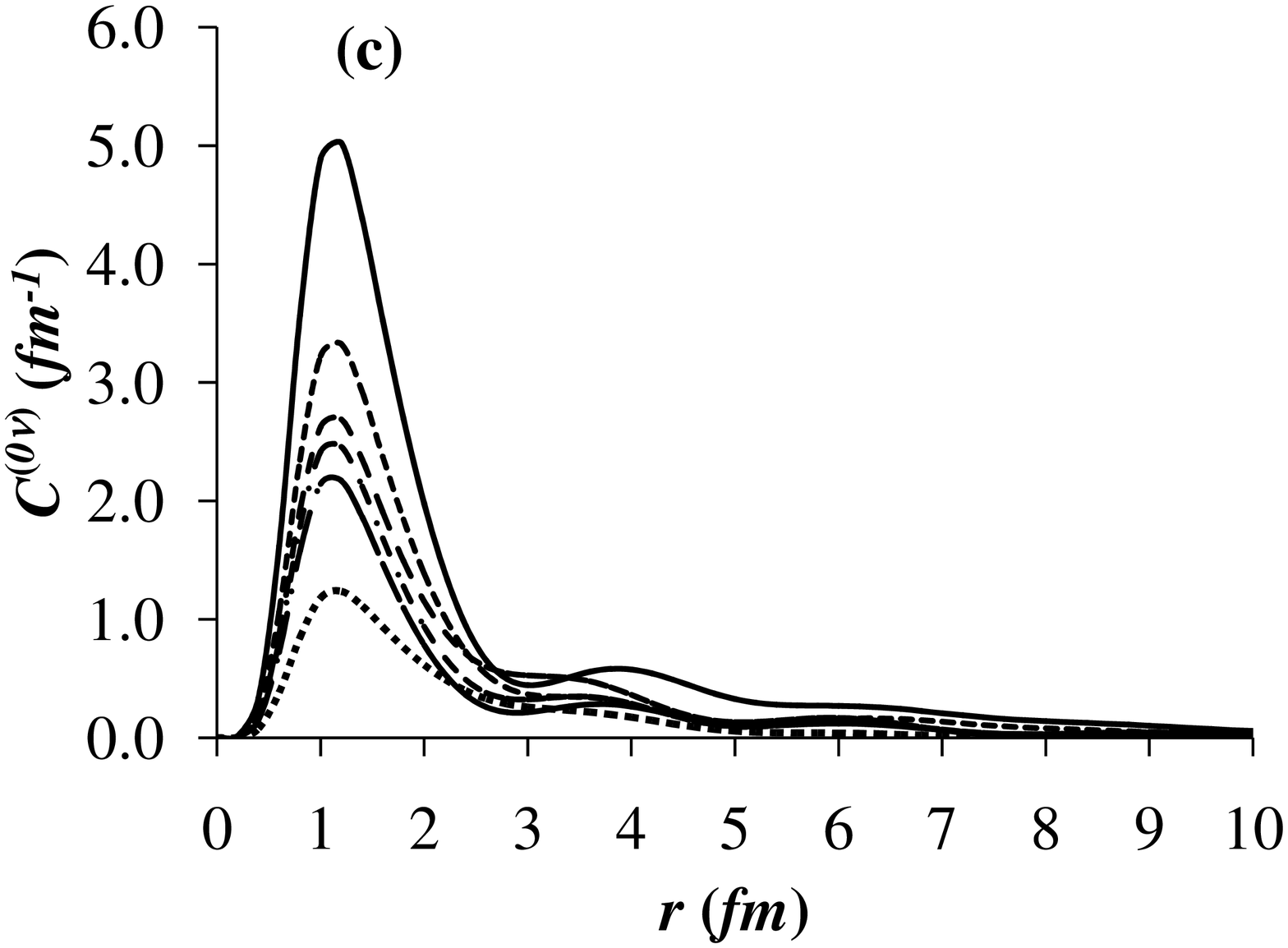} &
\includegraphics [scale=0.33]{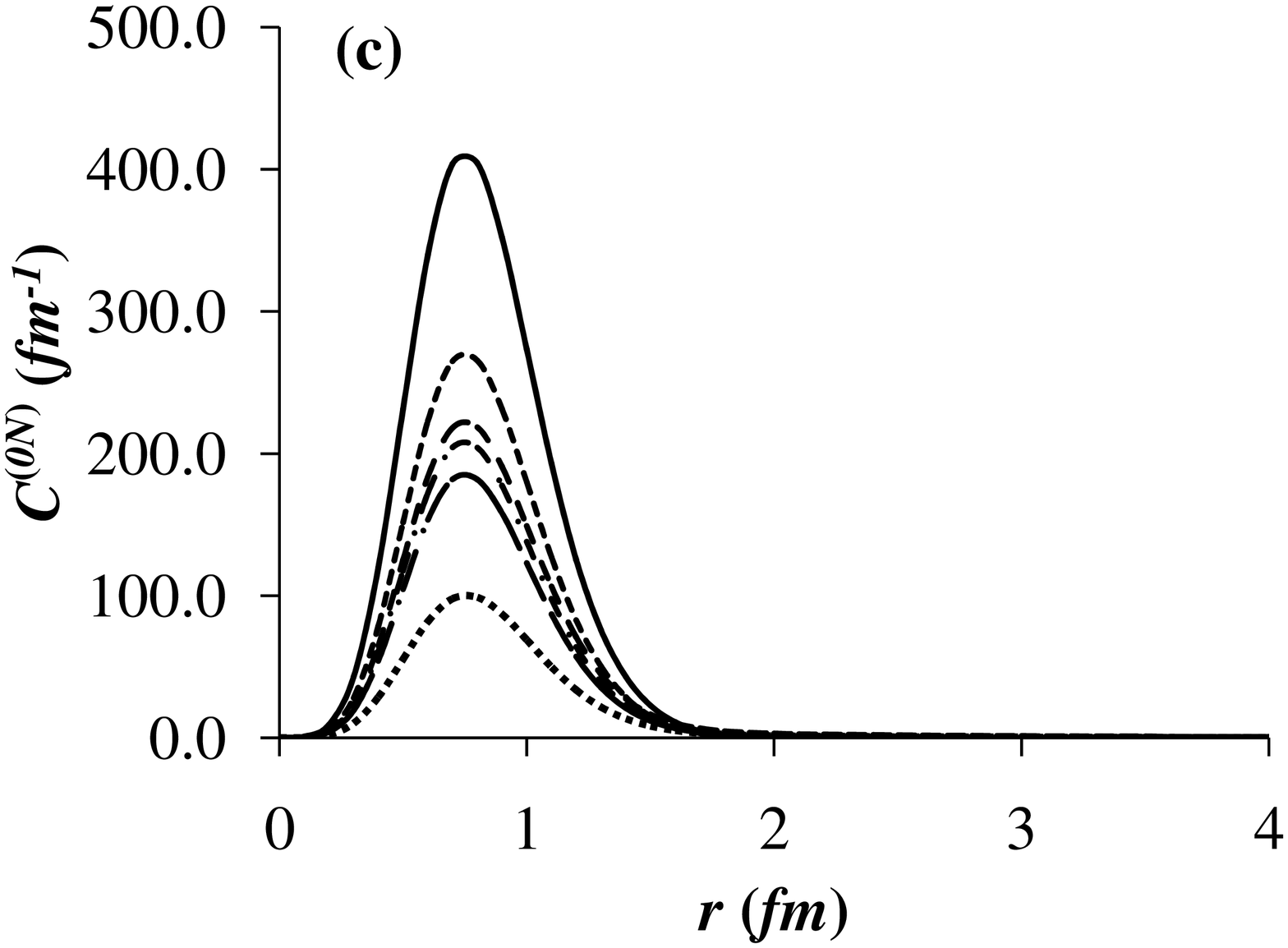} \\
\includegraphics [scale=0.33]{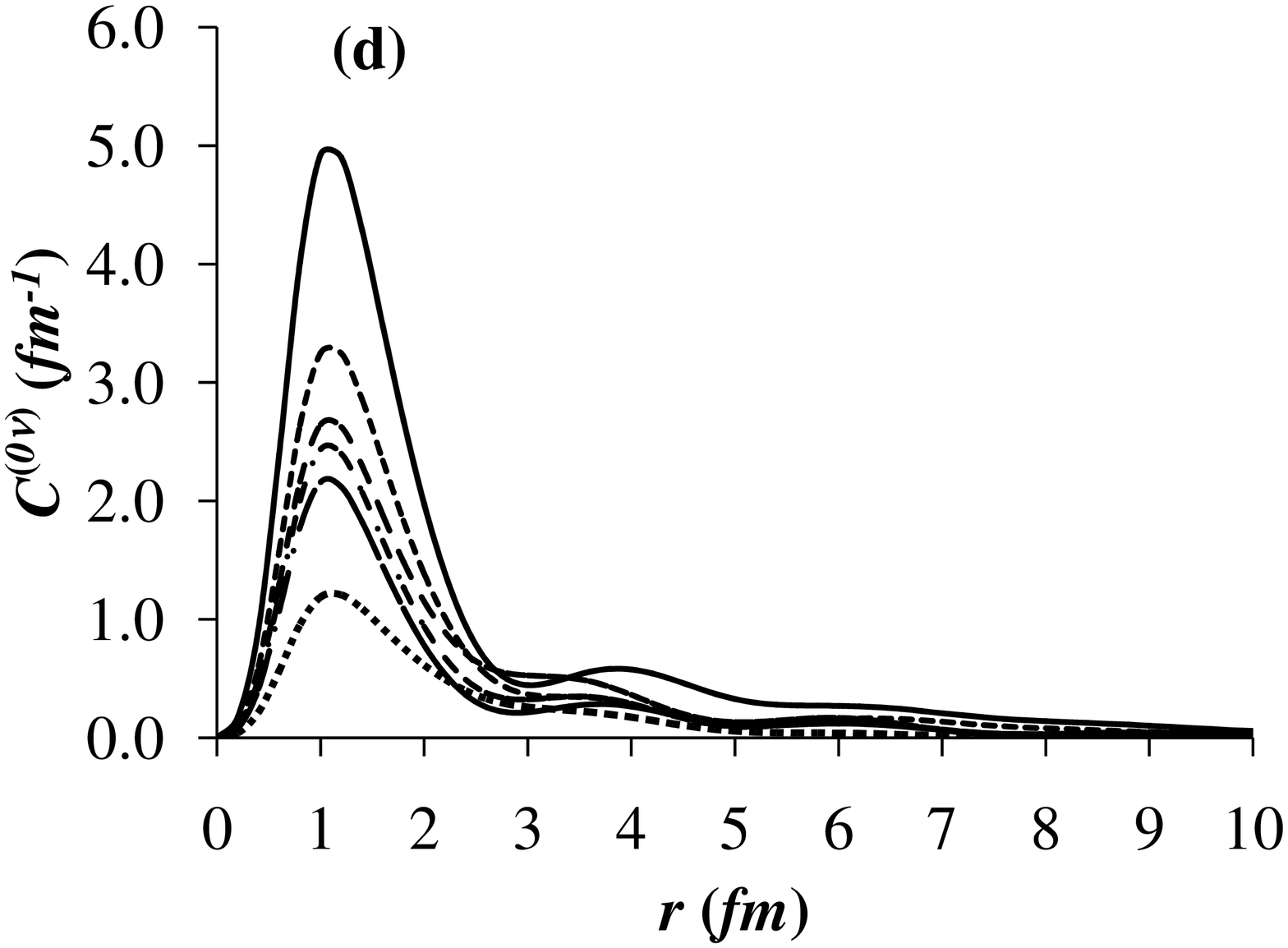} &
\includegraphics [scale=0.33]{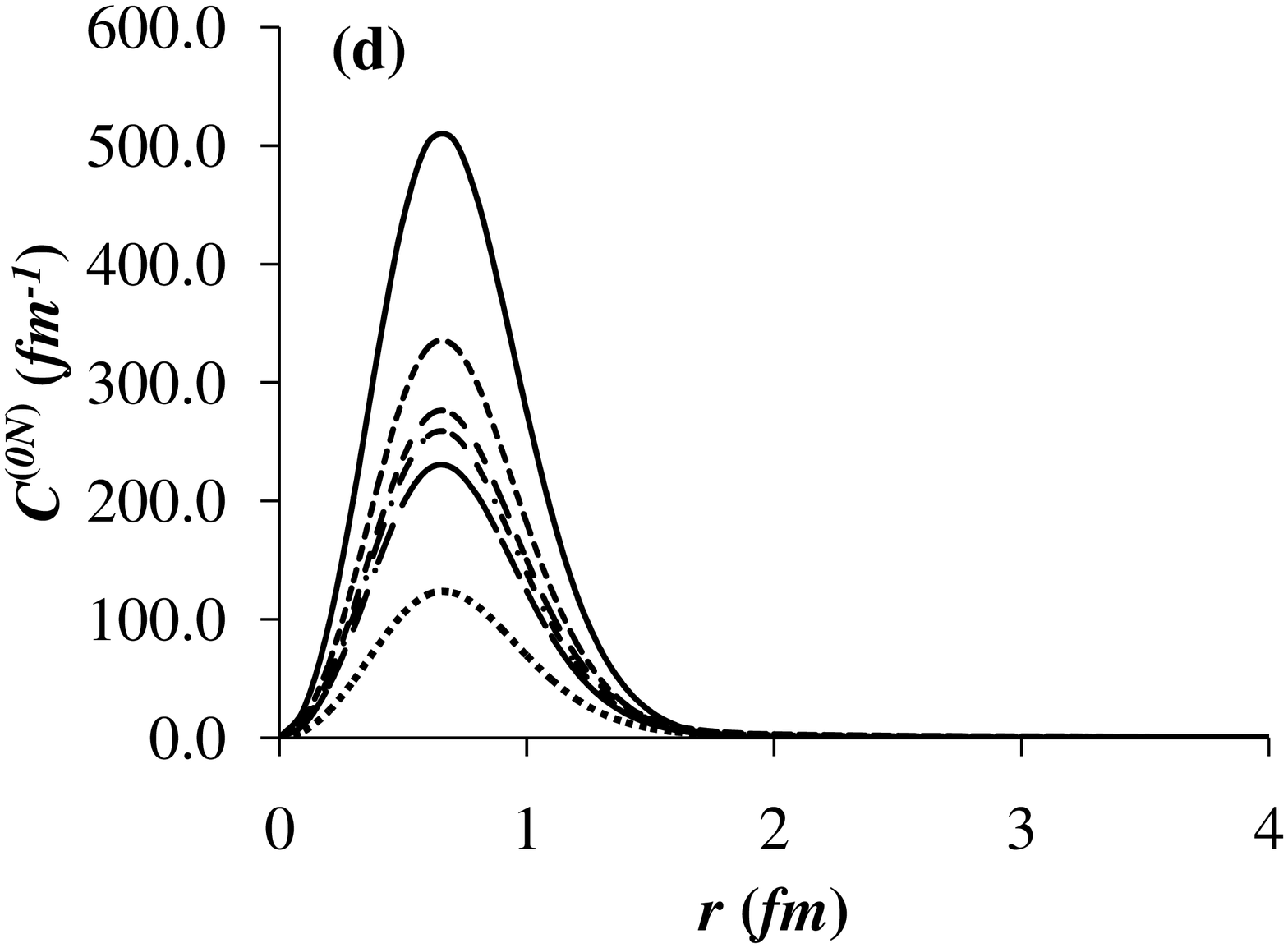} \\
\end{tabular}
\caption{Radial dependence of $C^{(0\nu)}(r)$ (left )and $C^{(0N)}(r)$ (right) for the
$\left( \beta ^{+}\beta ^{+} \right) _{0\nu }$ and
$\left( \varepsilon \beta ^{+}\right) _{0\nu }$ modes of
$^{96}$Ru, $^{102}$Pd, $^{106}$Cd, $^{124}$Xe, $^{130}$Ba and $^{156}$Dy
isotopes. In this Fig., (a), (b), (c) and (d) correspond to F, F+SRC1,
F+SRC2 and F+SRC3, respectively.}
\label{fig4}
\end{figure*}

Presently, we study the radial dependence of $C^{\left( 0\nu \right) }$
as well as $C^{\left( 0N \right) }$ for $(\beta ^{+}\beta ^{+})_{0\nu }$ and 
$(\varepsilon \beta
^{+})_{0\nu }$ modes of $^{96}$Ru, $^{102}$Pd, $^{106}$Cd, $^{124}$Xe, $%
^{130}$Ba and $^{156}$Dy isotopes in four cases, namely F, F+SRC1, F+SRC2 and
F+SRC3. In Fig.~\ref{fig2} we plot the radial dependence of $C^{\left( 0\nu \right) }$
and $C^{\left( 0N \right) }$ for $^{106}$Cd, employing the $PQQ1$
parameterization of the effective two body interaction, for four combinations of 
FNS and SRC. In Fig. 3, the radial evolution of $C^{\left( 0\nu \right) }$ and $C^{\left( 0N \right) }$
are displayed together for the six nuclei under study, for the four combinations of  FNS and SRC. 

In the case of light Majorana neutrino exchange, it is noticed that the 
$C^{\left( 0\nu \right) }$ are peaked at $r=1.0$ fm for finite size nucleons and
the addition of SRC1 and SRC2 shifts the peak to 1.25 fm. However, the position of 
the peak remains unchanged at $r=1.0$ fm with the inclusion of SRC3. The radial distributions of
$C^{\left( 0\nu \right) }$ extends up to 10 fm although the maximum contribution to 
$M^{\left( 0\nu \right) }$ results from the distribution up to 3 fm.
In the case of heavy Majorana neutrino exchange, the $C^{\left(0N \right) }$ are peaked 
at $r\approx 0.5$ fm in the case of FNS, and with the addition of SRC1 and SRC2, 
the peak shifts to about 0.8 fm, and to 0.7 fm for SRC3. 
The radial distributions of $C^{\left( 0N \right) }$ extend up to 2 fm and
the total distribution contributes to the evolution of $M^{\left( 0N \right) }$.  
Remarkably, the above observations also remain valid with the other three parameterizations 
of the effective two-body interaction.

\begin{table*} [htbp]
\caption{Average NTMEs $\overline{M}^{(K)}$ and uncertainties $\Delta
\overline{M}^{(K)}$ for $\left( \beta ^{+}\beta ^{+} \right) _{0\nu }$ and 
$\left( \varepsilon \beta ^{+}\right) _{0\nu }$ modes
of $^{96}$Ru, $^{102}$Pd, $^{106}$Cd, $^{124}$Xe, $^{130}$Ba and $^{156}$Dy 
isotopes. Both bare and quenched values of $g_{A}$ are considered. Case I
and Case II denote calculations with and without SRC1, respectively. In column 9, (l) and (s)
 denote large and small basis, respectively.}
\label{tab5}
\begin{tabular}{cccccccccccccccccccc}
\hline\hline
Nuclei~~& $g_{A}$ &~~  & \multicolumn{11}{c}{Light neutrino exchange} &~~~~ &
\multicolumn{5}{c}{Heavy neutrino exchange} \\ \cline{4-14}\cline{16-20}
&  &  & \multicolumn{2}{c}{Case I} &~~  & \multicolumn{2}{c}{Case II} &~~  & 
QRPA & QRPA &  & SQRPA & MCM &  & \multicolumn{2}{c}{Case I} &~~  &
\multicolumn{2}{c}{Case II} \\
\cline{4-5}\cline{7-8}\cline{16-17}\cline{19-20}
&  &  & $\overline{M}^{(0\nu )}$ & $\Delta \overline{M}^{(0\nu )}$ &  & $%
\overline{M}^{(0\nu )}$ & $\Delta \overline{M}^{(0\nu )}$ &  & \cite{hirs94}
& \cite{stau91} &  & \cite{stoi03} & \cite{suho03} &  & $%
\overline{M}^{(0N)}$ & $\Delta \overline{M}^{(0N)}$ &  & $\overline{M}^{(0N)}
$ & $\Delta \overline{M}^{(0N)}$ \\ \hline
$^{96}$Ru & 1.254 &  & 4.59 & 0.34 &  & 4.82 & 0.11 &  & 3.60 & 4.228  &  & 
& 2.383 &  & 148.13 & 50.27 &  & 178.29 & 29.19 \\
& 1.0 &  & 5.13 & 0.40 &  & 5.39 & 0.13 &  &  &  &  &  &  &  & 165.91 & 59.74
&  & 201.49 & 35.61 \\
&  &  &  &  &  &  &  &  &  &  &  &  &  &  &  &  &  &  &  \\
$^{102}$Pd & 1.254 &  & 4.71 & 0.60 &  & 4.97 & 0.50 &  &  &  &  &  &  &  &
160.83 & 58.13 &  & 195.01 & 35.86 \\
& 1.0 &  & 5.34 & 0.71 &  & 5.63 & 0.59 &  &  &  &  &  &  &  & 181.01 & 69.02
&  & 221.36 & 43.39 \\
&  &  &  &  &  &  &  &  &  &  &  &  &  &  &  &  &  &  &  \\
$^{106}$Cd & 1.254 &  & 7.57 & 0.89 &  & 7.97 & 0.72 &  & 4.56 & 4.778 &
& 7.85(l) & 3.394 &  & 249.89 & 89.73 &  & 302.93 & 54.54 \\
& 1.0 &  & 8.52 & 1.04 &  & 8.98 & 0.84 &  &  &  &  & 8.17(s)  &  &  & 281.22 &
106.57 &  & 343.82 & 66.13 \\
&  &  &  &  &  &  &  &  &  &  &  &  &  &  &  &  &  &  &  \\
$^{124}$Xe & 1.254 &  & 3.50 & 0.42 &  & 3.69 & 0.32 &  & 5.27 & 2.975  &  & 
& 8.301 &  & 124.84 & 44.75 &  & 151.45 & 26.71 \\
& 1.0 &  & 3.96 & 0.49 &  & 4.19 & 0.37 &  &  &  &  &  &  &  & 140.70 & 53.16
&  & 172.11 & 32.44 \\
&  &  &  &  &  &  &  &  &  &  &  &  &  &  &  &  &  &  &  \\
$^{130}$Ba & 1.254 &  & 2.60 & 0.80 &  & 2.75 & 0.82 &  & 5.52 & 5.579 &  & 
& 5.130 &  & 97.35 & 41.65 &  & 118.11 & 34.03 \\
& 1.0 &  & 2.94 & 0.90 &  & 3.12 & 0.92 &  &  &  &  &  &  &  & 109.75 & 48.71
&  & 134.24 & 39.56 \\
&  &  &  &  &  &  &  &  &  &  &  &  &  &  &  &  &  &  &  \\
$^{156}$Dy & 1.254 &  & 2.22 & 0.31 &  & 2.33 & 0.29 &  &  &  &  &  &  &  &
72.96 & 26.36 &  & 87.78 & 18.02 \\
& 1.0 &  & 2.50 & 0.36 &  & 2.63 & 0.33 &  &  &  &  &  &  &  & 81.66 & 31.12
&  & 99.13 & 21.41 \\ \hline\hline
\end{tabular}
\end{table*}

\begin{table*} [htbp]
\caption{Limits on the effective mass of light $<m_\nu>$ and heavy $<M_N>$
Majorana neutrinos for the $\left( \beta ^{+}\beta ^{+} \right) _{0\nu }$ and
$\left( \varepsilon \beta ^{+}\right) _{0\nu }$ modes
of $^{96}$Ru, $^{102}$Pd, $^{106}$Cd, $^{124}$Xe, $^{130}$Ba and $^{156}$Dy
isotopes.}
\label{tab6}
\begin{tabular}{cccccccccccccc}
\hline\hline
$e^{+}\beta \beta $ &~~~~~~  & \multicolumn{2}{c}{$T_{1/2}^{0\nu }$ (yr)} &
~~~~~ Ref.~~~~~ &
$g_{A}$ &~~~~~  & \multicolumn{3}{c}{$\left\langle m_{\nu }\right\rangle $ (eV)}
&~~~~~~  & \multicolumn{3}{c}{$\left\langle M_{N}\right\rangle $ (GeV)} \\
\cline{3-4}\cline{8-10}\cline{12-14}
emitters &  & $\beta ^{+}\beta ^{+}$ &~~~~~ $\varepsilon \beta ^{+}$ &  &  &  & $\beta ^{+}\beta ^{+}$ &~~~~  & $\varepsilon \beta ^{+}$ &  & $\beta ^{+}\beta
^{+}$ &~~~~  & $\varepsilon \beta ^{+}$ \\ \hline
$^{96}$Ru &  & $>${\small 3.1}$\times ${\small 10}$^{16}$ &~~~ $>${\small 6.7}$%
\times ${\small 10}$^{16}$ & \cite{norm85} & \multicolumn{1}{l}{\small 1.254}
&  & {\small 4.02}$\times ${\small 10}$^{5}$ &  & {\small %
7.94}$\times ${\small 10}$^{4}$ &  & {\small 4.41}$%
\times ${\small 10} &  & {\small 2.23}$%
\times ${\small 10}$^{2}$ \\
&  &  &  &  & \multicolumn{1}{l}{\small 1.0} &  & {\small 5.66}$%
\times ${\small 10}$^{5}$ &  & {\small 1.12}$%
\times ${\small 10}$^{5}$ &  & {\small 3.17}$%
\times ${\small 10} &  & {\small 1.61}$%
\times ${\small 10}$^{2}$ \\
&  &  &  &  &  &  &  &  &  &  &  &  &  \\
$^{106}$Cd &  & $>${\small 1.2}$\times ${\small 10}$^{21}$ &~~~ $>${\small 2.2}$%
\times ${\small 10}$^{21}$ & \cite{bell12} & \multicolumn{1}{l}{\small 1.254}
&  & {\small 1.16}$\times ${\small 10}$^{3}$ &  & {\small %
2.27}$\times ${\small 10}$^{2}$ &  & {\small 1.57}$%
\times ${\small 10}$^{4}$ &  & {\small 8.04}$%
\times ${\small 10}$^{4}$ \\
&  &  &  &  & \multicolumn{1}{l}{\small 1.0} &  & {\small 1.62}$%
\times ${\small 10}$^{3}$ &  & {\small 3.16}$%
\times ${\small 10}$^{2}$ &  & {\small 1.13}$%
\times ${\small 10}$^{4}$ &  & {\small 5.80}$%
\times ${\small 10}$^{4}$ \\
&  &  &  &  &  &  &  &  &  &  &  &  &  \\
$^{124}$Xe &  & $>${\small 4.2}$\times ${\small 10}$^{17}$ &~~~ $>${\small 1.2}$%
\times ${\small 10}$^{18}$ & \cite{bara89} & \multicolumn{1}{l}{\small 1.254}
&  & {\small 1.22}$\times ${\small 10}$^{5}$ &  & {\small %
1.68}$\times ${\small 10}$^{4}$ &  & {\small 1.61}$%
\times ${\small 10}$^{2}$ &  & {\small 1.17}$%
\times ${\small 10}$^{3}$ \\
&  &  &  &  & \multicolumn{1}{l}{\small 1.0} &  & {\small 1.70}$%
\times ${\small 10}$^{5}$ &  & {\small 2.33}$%
\times ${\small 10}$^{4}$ &  & {\small 1.16}$%
\times ${\small 10}$^{2}$ &  & {\small 8.46}$%
\times ${\small 10}$^{2}$ \\
&  &  &  &  &  &  &  &  &  &  &  &  &  \\
$^{130}$Ba &  & $>${\small 4.0}$\times ${\small 10}$^{21}$ &~~~ $>${\small 4.0}$%
\times ${\small 10}$^{21}$ & \cite{bara96a} & \multicolumn{1}{l}{\small 1.254%
} &  & {\small 4.11}$\times ${\small 10}$^{3}$ &  & {\small %
4.19}$\times ${\small 10}$^{2}$ &  & {\small 5.01}$%
\times ${\small 10}$^{3}$ &  & {\small 4.91}$%
\times ${\small 10}$^{4}$ \\
\multicolumn{1}{l}{} &  &  &  &  & \multicolumn{1}{l}{\small 1.0} &  &
\multicolumn{1}{l}{{\small 5.70}$\times ${\small 10}$^{3}$}
&  & \multicolumn{1}{l}{{\small 5.82}$\times ${\small 10}$%
^{2}$} & \multicolumn{1}{l}{} & \multicolumn{1}{l}{{\small 3.62}$%
\times ${\small 10}$^{3}$} &  & \multicolumn{1}{l}{{\small %
3.55}$\times ${\small 10}$^{4}$} \\ \hline\hline
\end{tabular}
\end{table*}

\begin{table*}
\caption{Predicted half-lives, corresponding extracted effective mass of heavy
Majorana neutrino $<M_{N}>$, nuclear sensitivities 
$\xi ^{\left( 0\nu \right) }$ and $\xi ^{\left(0N\right) }$ due to 
exchange of light and heavy neutrino, respectively, for $%
\left( \beta ^{+}\beta ^{+}\right) _{0\nu }$ and $\left( \varepsilon \beta
^{+}\right) _{0\nu }$ modes of $^{96}$Ru, $^{102}$Pd, $^{106}$Cd, $^{124}$%
Xe, $^{130}$Ba and $^{156}$Dy isotopes.}
\label{tab7}
\begin{tabular}{cccccccccccccccccc}
\hline\hline
$e^{+}\beta \beta $ &~~~~~  & $g_{A}$ &~~~~  &  & 
\multicolumn{4}{c}{$T_{1/2}^{0\nu }$
$(\left\langle m_{\nu }\right\rangle =0.05$ eV$)$} &~~~~  & $\left\langle
M_{N}\right\rangle $ (GeV) & ~~~~ &  & \multicolumn{2}{c}{$\xi ^{\left( 0\nu
\right) }$} &  & \multicolumn{2}{c}{$\xi ^{\left( 0N\right) }$} \\
\cline{5-9}\cline{11-11}\cline{13-18}
emitters &  &  &  &  & $\beta ^{+}\beta ^{+}$ &~~~~  &  & 
$\varepsilon \beta ^{+}
$ &  & $\beta ^{+}\beta ^{+}/\varepsilon \beta ^{+}$ &  &  & $\beta
^{+}\beta ^{+}$ & ~~~~$\varepsilon \beta ^{+}$~~~~ &  & 
$\beta ^{+}\beta ^{+}$ & 
~~~~$\varepsilon \beta ^{+}$~~~~ \\ \hline
$^{96}$Ru &  & \multicolumn{1}{l}{\small 1.254} &  &  & {\small 2.01}$%
_{-0.09}^{+0.09}\times ${\small 10}$^{30}$ &  &  & {\small 1.69}$%
_{-0.07}^{+0.08}\times ${\small 10}$^{29}$ &  & {\small 3.55}$%
_{-0.58}^{+0.58}\times ${\small 10}$^{8}$ &  &  & {\small 0.721} & {\small %
2.486} &  & {\small 26.70} & {\small 92.02} \\
&  & \multicolumn{1}{l}{\small 1.0} &  &  & {\small 3.97}$%
_{-0.18}^{+0.19}\times ${\small 10}$^{30}$ &  &  & {\small 3.34}$%
_{-0.15}^{+0.16}\times ${\small 10}$^{29}$ &  & {\small 3.59}$%
_{-0.63}^{+0.63}\times ${\small 10}$^{8}$ &  &  & {\small 0.513} & {\small %
1.768} &  & {\small 19.19} & {\small 66.13} \\
&  & \multicolumn{1}{l}{} &  &  &  &  &  &  &  &  &  &  &  &  &  &  &  \\
$^{102}$Pd &  & \multicolumn{1}{l}{\small 1.254} &  &  & {\small -} &  &  &
{\small 7.04}$_{-1.23}^{+1.66}\times ${\small 10}$^{30}$ &  & {\small 3.76}$%
_{-0.69}^{+0.69}\times ${\small 10}$^{8\dagger }$ &  &  & {\small -} &
{\small 0.385} &  & {\small -} & {\small 15.11} \\
&  & \multicolumn{1}{l}{\small 1.0} &  &  & {\small -} &  &  & {\small 1.36}$%
_{-0.25}^{+0.34}\times ${\small 10}$^{31}$ &  & {\small 3.77}$%
_{-0.74}^{+0.74}\times ${\small 10}$^{8\dagger }$ &  &  &  & {\small 0.277}
&  &  & {\small 10.90} \\
&  & \multicolumn{1}{l}{} &  &  &  &  &  &  &  &  &  &  &  &  &  &  &  \\
$^{106}$Cd &  & \multicolumn{1}{l}{\small 1.254} &  &  & {\small 6.49}$%
_{-1.03}^{+1.34}\times ${\small 10}$^{29}$ &  &  & {\small 4.52}$%
_{-0.71}^{+0.94}\times ${\small 10}$^{28}$ &  & {\small 3.64}$%
_{-0.66}^{+0.66}\times ${\small 10}$^{8}$ &  &  & {\small 1.268} & {\small %
4.806} &  & {\small 48.20} & {\small 182.6} \\
&  & \multicolumn{1}{l}{\small 1.0} &  &  & {\small 1.27}$%
_{-0.21}^{+0.28}\times ${\small 10}$^{30}$ &  &  & {\small 8.81}$%
_{-1.45}^{+1.92}\times ${\small 10}$^{28}$ &  & {\small 3.67}$%
_{-0.71}^{+0.71}\times ${\small 10}$^{8}$ &  &  & {\small 0.908} & {\small %
3.442} &  & {\small 34.79} & {\small 131.8} \\
&  & \multicolumn{1}{l}{} &  &  &  &  &  &  &  &  &  &  &  &  &  &  &  \\
$^{124}$Xe &  & \multicolumn{1}{l}{\small 1.254} &  &  & {\small 2.51}$%
_{-0.39}^{+0.51}\times ${\small 10}$^{30}$ &  &  & {\small 1.35}$%
_{-0.21}^{+0.27}\times ${\small 10}$^{29}$ &  & {\small 3.93}$%
_{-0.69}^{+0.69}\times ${\small 10}$^{8}$ &  &  & {\small 0.645} & {\small %
2.777} &  & {\small 26.44} & {\small 113.9} \\
&  & \multicolumn{1}{l}{\small 1.0} &  &  & {\small 4.84}$%
_{-0.76}^{+1.00}\times ${\small 10}$^{30}$ &  &  & {\small 2.61}$%
_{-0.41}^{+0.54}\times ${\small 10}$^{29}$ &  & {\small 3.94}$%
_{-0.74}^{+0.74}\times ${\small 10}$^{8}$ &  &  & {\small 0.465} & {\small %
2.001} &  & {\small 19.11} & {\small 82.30} \\
&  & \multicolumn{1}{l}{} &  &  &  &  &  &  &  &  &  &  &  &  &  &  &  \\
$^{130}$Ba &  & \multicolumn{1}{l}{\small 1.254} &  &  & {\small 2.70}$%
_{-1.09}^{+2.76}\times ${\small 10}$^{31}$ &  &  & {\small 2.82}$%
_{-1.14}^{+2.88}\times ${\small 10}$^{29}$ &  & {\small 4.12}$%
_{-1.19}^{+1.19}\times ${\small 10}$^{8}$ &  &  & {\small 0.197} & {\small %
1.926} &  & {\small 8.45} & {\small 82.68} \\
&  & \multicolumn{1}{l}{\small 1.0} &  &  & {\small 5.20}$%
_{-2.11}^{+5.31}\times ${\small 10}$^{31}$ &  &  & {\small 5.42}$%
_{-2.20}^{+5.54}\times ${\small 10}$^{29}$ &  & {\small 4.13}$%
_{-1.22}^{+1.22}\times ${\small 10}$^{8}$ &  &  & {\small 0.142} & {\small %
1.387} &  & {\small 6.11} & {\small 59.76} \\
&  & \multicolumn{1}{l}{} &  &  &  &  &  &  &  &  &  &  &  &  &  &  &  \\
$^{156}$Dy &  & \multicolumn{1}{l}{\small 1.254} &  &  & {\small -} &  &  &
{\small 5.94}$_{-1.23}^{+1.79}\times ${\small 10}$^{29}$ &  & {\small 3.62}$%
_{-0.74}^{+0.74}\times ${\small 10}$^{8\dagger }$ &  &  & {\small -} &
{\small 1.327} &  & {\small -} & {\small 50.04} \\
&  & \multicolumn{1}{l}{\small 1.0} &  &  & {\small -} &  &  & {\small 1.15}$%
_{-0.24}^{+0.35}\times ${\small 10}$^{30}$ &  & {\small 3.61}$%
_{-0.78}^{+0.78}\times ${\small 10}$^{8\dagger }$ &  &  &  & {\small 0.953}
&  &  & {\small 35.94} \\ \hline\hline
\end{tabular}
\footnote{$\dagger$ denotes $(\varepsilon \beta ^{+})_{0\nu}$ mode only.} 
\end{table*}

\subsection{Uncertainties in nuclear transition matrix elements and nuclear
sensitivity}

The uncertainties associated with the NTMEs $M^{(0\nu )}$ and $M^{(0N
)}$ for $(\beta ^{+}\beta ^{+})_{0\nu }$ and $(\varepsilon \beta ^{+})_{0\nu
}$ modes of $^{96}$Ru, $^{102}$Pd, $^{106}$Cd, $^{124}$Xe, $^{130}$Ba and $%
^{156}$Dy isotopes due to the exchange of light and heavy neutrinos,
respectively are evaluated by calculating the mean and standard
deviation given by 
\begin{equation}
\overline{M}^{(K )}=\frac{\sum_{i=1}^{N}M_{i}^{(K )}}{N}
\end{equation}
and 
\begin{equation}
\Delta \overline{M}^{(K )}=\frac{1}{\sqrt{N-1}}\left[
\sum_{i=1}^{N}\left( \overline{M}^{(K )}-M_{i}^{(K)}\right)
^{2}\right] ^{1/2}.
\end{equation}
The twelve NTMEs due to the exchange of light as well as heavy Majorana neutrinos 
listed in the three columns 4--6 and 11--13 (F+S) of Table~\ref{tab1} are employed 
in this statistical analysis for the bare and quenched
values of axial vector coupling constant $g_{A}=1.254$ and $g_{A}=1.0$,
respectively. Further, the effect due to the Miller-Spenser parameterization
of Jastrow type of SRC is estimated by evaluating the same mean $%
\overline{M}^{(K)}$ and their standard deviations $\Delta \overline{M}%
^{(K)}$ for eight NTMEs calculated using SRC2 and SRC3
parameterizations. In Table~\ref{tab5}, we display the calculated averages and
their variances along with all the available theoretical results in other models
for $^{96}$Ru, $^{102}$Pd, $^{106}$Cd, $^{124}$Xe, $^{130}$%
Ba and $^{156}$Dy isotopes.

In the case of light Majorana neutrino exchange, it is observed that the
uncertainties $\Delta \overline{M}^{(0\nu )}$ but for $^{130}$Ba are about
7\%--14\% and the exclusion of NTMEs $M^{\left( 0\nu \right) }$ calculated
with the Miller-Spencer parameterization of Jastrow SRC, reduces the uncertainties
to 2\%--12\% for both $g_{A}=1.254$ and $g_{A}=1.0$. Pathologically, the
uncertainty $\Delta \overline{M}^{(0\nu )}\approx 30\%$ in the case of $%
^{130}$Ba, remain unaltered due to the large effects of $PQQHH2$
parameterization. The estimated uncertainties $\Delta \overline{M}^{(0\nu )}$
but for $^{130}$Ba isotope in the heavy Majorana neutrino mass mechanism,
are about 35\% for $g_{A}=1.254$ and $g_{A}=1.0$. Estimation of
uncertainties for eight NTMEs $\overline{M}^{(0N )}$ calculated using
the SRC2 and SRC3 parameterizations again reveal that the $\Delta \overline{M}^{(0N )}$ 
are reduced to 16\%--21\% due to the exclusion of SRC1. In $%
^{130}$Ba isotope, the same pathological behaviour is noticed.

In the QRPA calculations of Hirsch \textit{et al.} \cite{hirs94} and 
Staudt \textit{et al. }\cite{stau91}, the NTMEs $M^{\left( 0\nu \right)
} $ are almost identical but for $^{124}$Xe, in which the difference is
approximately by a factor of 1.8. Stoica \textit{et al.} \cite{stoi03} have
used SQRPA model with two model spaces, namely small basis (oscillator
shells of $3\hbar \omega -5\hbar \omega +i_{13/2}$ orbit) and a large basis
(oscillator shells of $2\hbar \omega -5\hbar \omega +i_{13/2}$ orbit). They
used the same SPEs as those of Hirsch \textit{et al. }and an effective two-body 
interaction derived from the Bonn-A potential. The NTMEs calculated in
the SQRPA \cite{stoi03} do not depend much on the model space and differ by
a factor of 1.8 approximately from those of Hirsch \textit{et al.} \cite
{hirs94}.
 
There are no available theoretical results and experimental half-life 
limits for the $^{102}$Pd and $^{156}$Dy isotopes. The extracted limits on the 
effective light neutrino mass $<m_{\nu }>$ as
well as heavy neutrino mass $<M_{N}>$ using the phase space factors given in 
Ref. \cite{rath09} and presently available experimental
limits on observed half-lives of $\left( \beta ^{+}\beta ^{+}\right) _{0\nu
} $ and $\left( \varepsilon \beta ^{+}\right) _{0\nu }$ modes are presented
in Table~\ref{tab6}. The extracted limits on $<m_{\nu }>$ and $<M_{N}>$ are not so
much stringent as in the case of $\left( \beta ^{-}\beta ^{-}\right) _{0\nu
} $ decay. Moreover, better limits are obtained in the case of $\left(
\varepsilon \beta ^{+}\right) _{0\nu }$ mode even for equal limits on
half-lives of $\left( \beta ^{+}\beta ^{+}\right) _{0\nu }$ and $\left(
\varepsilon \beta ^{+}\right) _{0\nu }$ modes. The best obtained limits for $%
^{106}$Cd isotope are $<m_{\nu }>$ $<1.16\times 10^{3}$ $%
(2.27\times 10^{2})$ eV and $<M_{N}>>1.57%
\times 10^{4}$ $(8.04\times 10^{4})$ GeV in case of $\left(
\beta ^{+}\beta ^{+}\right) _{0\nu }$ and $\left( \varepsilon \beta
^{+}\right) _{0\nu }$ modes, respectively. 
In the case of $\left( \beta ^{+}\beta ^{+}\right) _{0\nu }$ and $%
\left( \varepsilon \beta ^{+}\right) _{0\nu }$ modes, the extracted limits
on the effective neutrino masses $<m_{\nu }>$ and $<M_{N}>$ are not
stringent enough and hence, we calculate half-lives of these modes to be
useful in the design of future experimental setups. The half-lives of $%
\left( \beta ^{+}\beta ^{+}\right) _{0\nu }$ and $\left( \varepsilon \beta
^{+}\right) _{0\nu }$ modes for $<m_{\nu }>=50$ $meV$ are calculated and
extracted corresponding limits on heavy neutrino mass, $<M_{N}>$, are given in
the same Table~\ref{tab7}.

In the absence of stringent limits on the effective neutrino masses $<m_{\nu
}>$ and $<M_{N}>$, it is useful to calculate the nuclear sensitivity, defined 
as \cite{simk99}
\begin{equation}
\xi ^{\left( K\right) }=10^{8}\sqrt{G_{01}}\left| M^{(K)}\right|
\end{equation}
where $K$ stands for $0\nu $ or $0N$ mode and an arbitrary normalization
factor 10$^{8}$ is introduced so that the nuclear sensitivity turns out to
be order of unity. 

It is observed that in general, nuclear sensitivities for
$\left( \varepsilon \beta ^{+}\right) _{0\nu }$ mode are larger than those
of $\left( \beta ^{+}\beta ^{+}\right) _{0\nu }$ mode. Further, the nuclear
sensitivities for $\left( \beta ^{+}\beta ^{+}\right) _{0\nu }$ and 
$\left( \varepsilon \beta ^{+}\right) _{0\nu }$ modes 
of $^{106}$Cd, $^{96}$Ru ($^{124}$Xe), $^{124}$Xe ($^{96}$Ru), $^{130}$Ba, 
$^{156}$Dy and $^{102}$Pd isotopes, respectively, are in the decreasing order 
of their magnitudes.

\section{CONCLUSIONS}

We have calculated sets of twelve NTMEs $M^{\left( 0\nu \right) }$
and $M^{\left( 0N \right) }$ for $\left( \beta ^{+}\beta ^{+}\right)
_{0\nu }$ and $\left( \varepsilon \beta ^{+}\right) _{0\nu }$ modes of
$^{96}$Ru, $^{102}$Pd, $^{106}$Cd, $^{124}$Xe, $^{130}$Ba and $^{156}$Dy
isotopes by employing the PHFB model with four different parameterizations
of the pairing plus multipolar type of effective two body interaction and
three different parameterizations of the short range correlations.
To estimate statistically the uncertainties in NTMEs, mean and standard deviations of
sets of twelve NTMEs $M^{\left( 0\nu \right) }$ and $M^{\left( 0N \right) }$
calculated with dipole form factor and short range correlations are employed 
for both $g_{A}=1.254$ and $g_{A}=1.0$. 
It is observed that the largest standard deviation turns out to be around 30\% in
the case of $^{130}$Ba isotope due to the dominant contribution of
deformation in $PQQHH2$ parameterization. But for $^{130}$Ba, the maximum
uncertainty in NTMEs $M^{\left( 0\nu \right) }$ is around 14\%, which
becomes smaller by 2\% excluding the NTMEs calculated with SRC1 in the case
of $^{156}$Dy isotope. The uncertainties in $M^{\left( 0N \right) }$
due to the exchange of heavy Majorana neutrino is about 35\%. Exclusion of
NTMEs calculated with SRC1, reduced the uncertainties by 14\%--19\%.

\begin{acknowledgments}
This work is partially supported by the Council of Scientific and Industrial
Research (CSIR), India vide sanction No. 03(1216)/12/EMR-II, Indo-Italian 
Collaboration DST-MAE project via grant no. INT/Italy/P-7/2012 (ER), Consejo 
Nacional de Ciencia y Tecnolog\'{i}a (Conacyt)-M\'{e}xico, 
and Direcci\'{o}n General de Asuntos del Personal Acad\'{e}mico,
Universidad Nacional Aut\'{o}noma de M\'{e}xico (DGAPA-UNAM) project IN103212.
\end{acknowledgments}

\end{document}